\documentclass[aps,prl,twocolumn,amsmath,amssymb,a4paper]{revtex4-1}

\usepackage{graphicx}
\usepackage{mathtools}
\usepackage{color}
\usepackage[normalem]{ulem}
\usepackage{tabularx}
\DeclarePairedDelimiter{\norm}{\lVert}{\rVert}
\usepackage{enumerate}

\begin{document}

\title{Designing patchy interactions to self-assemble arbitrary structures}
\date{\today}
\author{Flavio Romano$^{1,2}$}
\author{John Russo$^{1,3}$}
\author{Luk\'{a}\v{s} Kroc$^4$}
\author{Petr \v{S}ulc$^{1,4,}$}
\email[Corresponding author: ]{psulc@asu.edu}
\affiliation{
$^1$Dipartimento di Scienze Molecolari e Nanosistemi,
Universit\`{a} Ca' Foscari di Venezia
Campus Scientifico, Edificio Alfa,
via Torino 155, 30170 Venezia Mestre, Italy\\
$^2$European Centre for Living Technology (ECLT) Ca' Bottacin, 3911 Dorsoduro Calle Crosera, 30123 Venice, Italy \\
$^3$Dipartimento di Fisica, Sapienza Universit\`{a} di Roma, P.le Aldo Moro 5, 00185 Rome, Italy\\
$^4$School of Molecular Sciences and Center for Molecular Design and Biomimetics, The Biodesign Institute, Arizona State University, 1001 South McAllister Avenue, Tempe, Arizona 85281, USA \\
}

\begin{abstract}
One of the fundamental goals of nanotechnology is to exploit selective and directional interactions between molecules to design particles that self-assemble into desired structures, from capsids, to nano-clusters, to fully formed crystals with target properties (e.g. optical, mechanical, etc.). Here we provide a general framework which transforms the inverse problem of self-assembly of colloidal crystals into a Boolean satisfiability problem for which solutions can be found numerically. Given a reference structure and the desired number of components, our approach produces designs for which the target structure is an energy minimum, and also allows to exclude solutions that correspond to competing structures. We demonstrate the effectiveness of our approach by designing model particles that spontaneously nucleate milestone structures such as the cubic diamond, the pyrochlore and the clathrate lattices.
\end{abstract}

\maketitle

Self-assembly is a broad category of processes by which elementary components organise themselves into ordered structures~\cite{whitelam2015statistical}.
Inspired by its ubiquity in biology, nanotechnology has long looked at self-assembly as the most promising avenue for the bottom-up realization of structures with specific properties. 
Successful experimental examples include two-dimensional lattices~\cite{Chen2011}, fully three-dimensional crystals~\cite{chen1991synthesis,douglas2009self}, and polyhedral shells~\cite{takeda1999nanometre,bhatia2009icosahedral,liu2011supramolecular}. On the molecular scale, perhaps the most successful results were obtained using DNA nanotechnology, where DNA sequences are designed so that they form the maximum number of base pairs only by self-assembling into the desired target 2D or 3D shape, e.g.~DNA origami \cite{Rothemund2006}. Very recently, DNA origami have been crystallized into three dimensional superlattices \cite{liu2016diamond,zhang20183d}.
At the colloidal scale, promising strategies for self-assembly include DNA-functionalized particles and patchy particles. In the former case, a mixture is obtained from colloids whose surface is randomly decorated with single strands of DNA such that particles of different types can selectively bind to each other. This strategy has led to the self-assembly of the double diamond (or B32) crystal~\cite{wang2017colloidal}. In the case of patchy particles, colloidal particles acquire anisotropic interactions either via their shape~\cite{van2013entropically} or via chemical patterning of their surface~\cite{zhang2004self,pawar2010fabrication,bianchi2011patchy,romano2011colloidal}. Hybrid solutions where patchy interactions are realized by attaching DNA sequences at well-defined positions have also been proposed~\cite{suzuki2009controlling,kim2011dna,wang2012colloids,feng2013dna}.

The experimental methodologies so far described, while successful, are system-specific and hard to generalize. In many cases, we lack a theoretical understanding of why certain structures have self-assembled from elementary building blocks. The search for the general principles behind the inverse self-assembly problem has attracted several theoretical investigations. Instead of predicting  which structures self-assemble out of specific building blocks, the inverse problem is concerned with designing building blocks that form a specific target. So far, two types of approaches have emerged: optimization algorithms and geometrical strategies. In optimization algorithms the pair potential is tuned to minimize the energy of a target structure~\cite{rechtsman2005optimized,marcotte2011optimized,marcotte2013designeddiamond,zhang2013probing,miskin2016turning,chen2018inverse,kumar2019inverse}.
While powerful and general, the major limitation of this approach is that the level of control over the shape of the pair-potential is in most cases far beyond current experimental possibilities.
The geometric approach to self-assembly instead uses specific interactions to match the geometrical properties of the target structure to kinetically guide the assembly process. The following interaction properties are usually tuned to match the target structure: shape~\cite{ducrot2017colloidal}, directionality~\cite{nelson2002toward,manoharan2003dense,zhang2005self,romano2014influence}, selective binding~\cite{halverson2013dna}, and torsional interactions between  neighbors~\cite{romano2012patterning,tracey2019programming}. 
Geometrical approaches allow experimentally realizable systems to self-assemble into specific structures, but the process of designing the potential is system-specific and requires ad-hoc solutions. An example of these limitations is the self-assembly of the colloidal diamond structure, which usually requires either torsional interactions~\cite{romano2012patterning,tracey2019programming} or hierarchical assembly~\cite{morphew2018programming,patra2018programmable,tracey2019programming,ma2019inverse} to avoid the formation of stacking faults. Importantly, some of these features lack a convincing experimental counterpart.

\begin{figure}[t]
    \centering
    \includegraphics[width=0.48\textwidth]{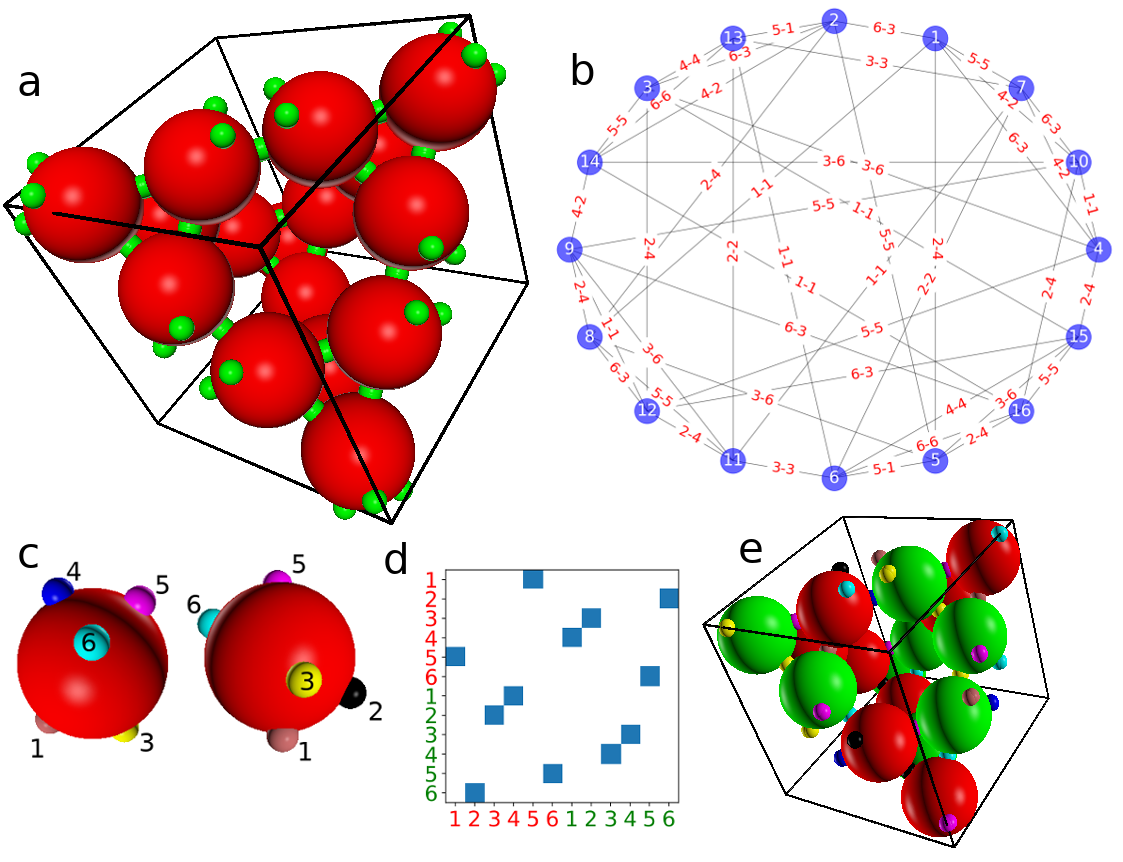}
    \caption{a) A schematic representation of the unit cell of a tetrastack lattice consisting of 16 positions (red spheres), each bound to its neighbors (using periodic boundary conditions) via numbered ``slots'', shown in green. b) A topology representing the unit lattice, showing each lattice position connected to 6 other positions (interacting slot numbers on each respective positions are shown as link labels). c) A PP with 6 patches with each patch colored differently. The PP can be positioned into the lattice so that its patches overlap with the green slots. There are 6 different orientations that allow to position a PP into the lattice position so that all patches overlap with the slots. d) The SAT solver assigns colors to each patch and designs the interaction matrix between the colors. In this particular solution for tetrastack crystal, there are 2 PP species (red and green) with each patch assigned its unique color. The interaction matrix shows which colors interact. e) The SAT solver assigns to each lattice positions a corresponding PP species and an orientation so that all patch interactions are satisfied. Patches that interact with each other are drawn using the same color for convenience.}
    \label{fig_sims}
\end{figure}

Here we formulate a general framework for designing self-assembling systems of patchy particles (PP) into any arbitrary structure, with the option to exclude the formation of competing structures that are identified in simulations.
We focus on designing PP systems that have geometric properties, such as the number and placement of the patches that reflect the local environment of the target lattice. To introduce selectivity in the model, we assign to each patch a ``color'' that encodes its binding properties and define $N_{\rm c}$ as the number of different patch colors. Binding is allowed only between patches that have compatible colors as specified by an interaction matrix (Fig.~\ref{fig_sims}d). We do not impose any torsional restrictions and all bonds have the same strength. While all particles have the same placement of the patches, we allow for the possibility of having $N_{\rm s}$ different PP ``species'', which are defined by the coloring of their patches. Relative concentrations are also free parameters. This system can be realised experimentally with patches based on single-stranded DNA~\cite{tian2020ordered} thanks to the selective binding of DNA sequences. To simplify sequence design, we impose that each patch is assigned a color that can only bind one other color (which can be the same in the case of self-complementarity). The goal is to determine the patch coloring for each PP species and color interaction matrix so that the PPs assemble into a desired structure.

The target structure is described by a unit cell, comprising $l \in [1,L]$ particles (Fig.~\ref{fig_sims}a). The unit cell can be a combination of two or more true unit cells of the target lattice.
The positions of the patches in the target lattice (slots) are labeled as $k \in [1,N_{\rm p}]$, from which a list of neighboring slots is computed (Fig.~\ref{fig_sims}b). Our designed PPs can be of different species $s \in [1,N_{\rm s}]$, and have $p \in [1,N_{\rm p}]$ patches on their surface which can take a color $c \in [1, N_{\rm c}]$. $o$ represents one of the $N_o$ possible orientations of each particle, and it is uniquely identified by a map between its patches and the patch slots they occupy. Not all mappings are possible, only those that can be reached by a physical rotation of the particle.

A brute-force search of all possible combinations of (i) patch color arrangements for all particle species, ${N_{\rm c} N_{\rm s} \choose N_{p}}$, (ii) rotations and symmetry operations of each particle, up to $N_{\rm p}^L!$, and (iii) interaction matrix between patch colors, ${ (N_{\rm c}+1)/2 \choose N_{\rm c}}$, becomes intractable with increasing $N_{\rm c}$, $N_{\rm s}$, $N_{\rm p}$ and $L$ even if they are relatively small.
Instead, and this a crucial contribution of this paper, we map the problem to a Boolean satisfiability problem (SAT), where recent algorithmic developments have dramatically advanced our ability to solve problems involving tens of thousands of variables and millions of constraints~\cite{een2005minisat,liang2018machine,papadimitriou1991selecting,jarvisalo2012international}.
We use a publicly available SAT solver~\cite{een2005minisat} that can find solutions to the design problems considered here in time ranging from few seconds up to one hour.

Mapping the particle design onto a SAT problem requires the definition of (i) binary variables $x_i$ that describe the PPs' patch coloring for each particle species and the color interaction matrix and (ii) binary clauses $C_j$ that represent the constraints that the variables need to satisfy, such as the ability to form all the bonds in the target lattice. Each binary clause contains a subset of the variables $x_i$ (or their negation $\neg x_i$) connected by an OR statement, and a solution is found whenever a combination of values of the $x_i$ satisfies all clauses at the same time. Formally, this corresponds to finding a set of variables $x_i$ such that $C_1 \land C_2 \land C_3 \land \ldots $ is true.
The SAT mapping can also be used to prove the impossibility of achieving some (desired or unwanted) binding pattern with a given combination of input parameters $N_{\rm s}$ and $N_{\rm c}$ by proving the absence of solutions to the associated SAT problem. This is crucial since it allows us to filter undesired binding patterns, which may represent competing global arrangements (i.e., another crystal form) or local arrangements (kinetic traps). 
\begin{figure*}[t]
    \centering
    \includegraphics[width=0.9\textwidth]{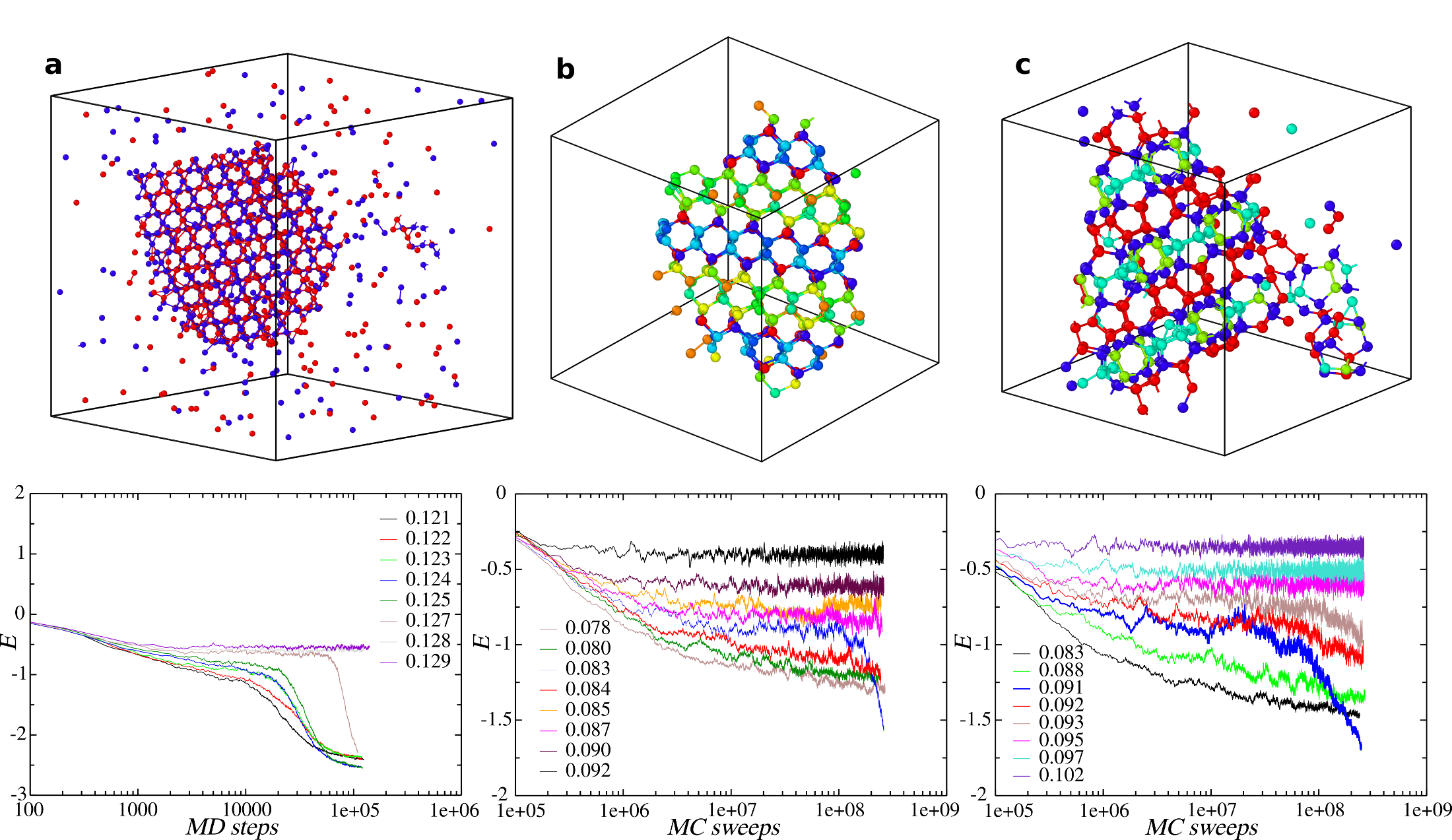}
    \caption{Overview of simulations of the assembly of \textbf{a)} tetrastack, \textbf{b)} diamond cubic, and \textbf{c)} clathrate Si34 lattices. Top panels show a snapshot of the final configuration of simulations that nucleated a crystal assembly, with each particle species colored differently. Patches are not shown for clarity. The bottom row shows the energy per particle over the course of the simulations at different temperatures, where nucleation events are signaled by sudden drops in $E$.  For tetrastack, we saw multiple independent nucleations for $T < 0.127$. Energy is reported in simulation energy unit $\epsilon^*$ and temperature is in $\epsilon^*/k_{\rm B}$.
    }
    \label{fig_simulation}
\end{figure*}

\begin{table}[b!]
	\centering
	\begin{tabular}{p{0.04\textwidth}p{0.12\textwidth}p{0.31\textwidth} } 
		\hline
		Id  & Clauses & Boolean expression \\ 
		\hline
		(i) & $C^{\rm int}_{c_i,c_j,c_k}$ & $ \neg x^{\rm int}_{c_i,c_j} \lor \neg x^{\rm int}_{c_i,c_k}$  \\
		(ii) & $C^{\rm pcol}_{s,p,c_k,c_l}$ & $\neg x^{pcol}_{s,p,c_k} \lor \neg x^{\rm pcol}_{s,p,c_l}$  \\ 
		(iii) & $ C^L_{l,s_i,o_i,s_j,o_j}$ & $\neg x^L_{l,s_i,o_i} \lor \neg x^L_{l,s_j,o_j}$ \\ 
		 (iv) & $C^{\rm lint}_{l_i,k_i,l_j,k_j,c_i,c_j}$ &  $(x^A_{l_i,k_i,c_i} \land  x^A_{l_j,k_j,c_j}) \implies x^{\rm int}_{c_i,c_j}$ \\
		(v) & $C^{\rm LS}_{l,s,o,c,k}$ &  $ x^L_{l,s,o} \implies \left( x^A_{l, k, c} \iff x^{\rm pcol}_{s, \phi_o(k), c} \right) $ \\
		(vi) & $C^{\rm all\,s.}_{s}$ & $  \bigvee_{\forall l , o } x^L_{l,s,o} $ \\
	    (vii)& $C^{\rm all\,c.}_{c}$ & $ \bigvee_{\forall s , p  } x^{\rm pcol}_{s,p,c}$ \\
		\hline
	\end{tabular}
	\caption{SAT clauses and variables.
	The color interaction is given by binary variables $x^{\rm int}_{c_i,c_j}$ which are 1 if color $c_i$ is compatible with color $c_j$ and 0 
	otherwise. The patch coloring for each PP species is described by binary variables $x^{\rm pcol}_{s,p,c}$ which are 1 if patch $p$ of species $s$ has color $c$ and 0 otherwise. 
	The arrangement of the particle species in the lattice is described by $x^{L}_{l,s,o}$ which is 1 if the position $l$ is occupied by a PP of species $s$ in the specific orientation $o$. The mapping $\phi_o(k) = p$ for a given orientation $o$ means that PP's patch $p$ overlaps with slot $k$ in a given lattice position. The variable $x^A_{l,k,c}$ is 1 if slot $k$ of lattice position $l$ is occupied by a patch with color $c$ and 0 otherwise. The clauses and variables are defined for all possible combinations of colors $c \in [1,N_c]$, patches $p \in [1,N_p]$, slots $k \in [1,N_p]$,  PP species $s \in [1,N_s]$, orientations $o \in [1,N_o]$, and lattice positions $l \in [1,L]$. Clauses $C^{\rm lint}$ are defined only for slots $k_i,k_j$ that are in contact in neighboring lattice positions $l_i, l_j$. 
	 For a given $s$, clause $C^{\rm all\,s.}_{s}$ is defined as a list of $x^L_{l,s,o}$ for all possible values of $l$ and $o$, joined by disjunctions. Clause $C^{\rm all\,c.}_{c}$ is defined analogously.
	}
	\label{table:sat}
\end{table}

Our design problem thus translates into a set of binary variables and clauses, as defined in Table~\ref{table:sat}.
In order, the clauses enforce that
(i) $C^{\rm int}$: each color is compatible with exactly one color,
(ii) $C^{\rm pcol}$: each patch is assigned exactly one color,
(iii) $C^L$: each lattice position is occupied by a single PP species with one assigned orientation,
(iv) $C^{\rm lint}$: colors of patches that interact in the target lattice can bind to each other according to the interaction matrix,
(v) $C^{\rm LS}$: the slots in each lattice position are set to have the color of the patch occupying them,
(vi) $C^{\rm all\,s.}_{s}$: all $N_{\rm s}$ particle species are used for the lattice assembly, (vii) $C^{\rm all\,c.}$: all $N_{\rm c}$ patch colors are used in the solution. 
The final SAT problem is a conjunction of all clauses (i)-(vii).
The conditions (vi) and (vii) are used to avoid getting trivial solutions such as having a single PP species with all patches colored by the same self-complementary color. It allows us to formulate the SAT problems for different combinations of $N_{\rm s}$ and $N_{\rm c}$ and see for which the solutions exist.

For the lattice design problems considered in this work, the number of binary variables ranges from about $10^3$ to $10^5$, and the number of clauses ranges from  approx.~$10^4$ to $10^7$, which we found to be within reach of a commonly used SAT solver~\cite{een2005minisat}. If a solution is found in terms of the binary variables $x$, it can be straightforwardly converted into human-readable form by listing the variables $x^{\rm pcol}_{s,p,c}$ and $x^{\rm int}_{c_i,c_j}$ that are 1, as their subscripts will specify respectively i) the color $c$ of patch $p$ in PP species $s$, ii) the compatible colors $c_i$ and $c_j$. 
Additionally, our framework allows the user to quickly check if a specific combination of PP species with a patch coloring and color interaction matrix can satisfy a given lattice geometry. We use clauses (i)-(iv) discussed above, and additionally add clauses that constrain the variables $x^{\rm pcol}$ and $x^{\rm int}$ accordingly for the set of PPs we want to check. If such a SAT problem is solvable, the indices of the variables $x^L_{l,s,o}$ that are 1 readily provide the particle species $s$ and orientation $o$ assigned to each lattice position $l$, allowing visualization of the lattice (Fig.~\ref{fig_sims}e).

To demonstrate the versatility of the SAT mapping approach for particle design, we selected three of the most challenging and sought after lattice geometries. After using our SAT solver to obtain PP species design and color interaction matrix, we run molecular simulations \cite{rovigatti2018simulate} and study the success and quality of the crystals obtained from homogenous nucleation. The SAT solver guarantees that the target structure is an energy minimum, but cannot say whether kinetic traps or other energy minima, both often associated with competing crystalline structures, are present along the self-assembly pathway.
If a competing structure is found in the molecular simulations, we can explicitly exclude it by redesigning the SAT problem by adding additional clauses or by discarding all generated solutions that can form the identified undesired structures. 
Thus we iteratively arrive at a design which self-assembles into the desired crystal through homogeneous nucleation. 
In contrast to previous solutions to this problem, we stress that these crystalline structures are being nucleated without introducing torsional interactions or hierarchical assembly. Moreover, the crystals are nucleated homogeneously, without the need for seeding or templating, and grow without stacking defects.

Our first target structure is the cubic tetrastack (TS) lattice (also known as pyrochlore) that together with the cubic diamond has been proposed for its omnidirectional photonic band gap for use as a photonic crystal~\cite{ngo2006tetrastack}. We adopt a design with 6 patches in the direction of the closest neighbours (Fig.~\ref{fig_sims}a,c). To mimic the possible experimental realization of the system using 3D DNA nanostructures \cite{veneziano2016designer}, with single-stranded DNA representing individual patches, we model the PPs as soft spheres with attractive point-patches (as described in Supp.~Mat.). Solutions can be found with $N_{\rm s} = 1, N_{\rm c} = 3$ but suffer from the geometric problem in which two particles can bind with two bonds at the same time, leading to alternative assemblies in the simulations. To avoid this we introduced an additional clause (defined in the Supp. Mat.) that requires that no pair of particles can bind through more than one bond.
This SAT problem has no solution for $N_{\rm s} = 1$. For $N_{\rm s} = 2$, we found solutions for $N_{\rm c} = 6, 8, 10,$ and $12$. We show the solution for $N_{\rm c} = 12$ in Fig.~\ref{fig_sims}d,e and its successful  nucleation in Fig.~\ref{fig_simulation}a. To simulate the assembly kinetics, we used simulations at a range of temperatures of a point-patch particle model~\cite{rovigatti2015comparison} (see the Supp. Mat.~for model and simulation details). The simulations were carried out with $2048$ particles at number density $0.1$ (corresponding to a volume fraction of $\approx 0.05$). This density was chosen to mimic the common experimental scenario of phase-separation-induced crystallization from a low-density solution. Fig.~\ref{fig_simulation}a shows a nucleation event: the top panel shows a snapshot of the nucleus, the bottom panel shows the time evolution of the energy for runs at different temperatures, where the nucleating trajectory is signaled by the sharp decrease in energy. 
We further obtained successful nucleation for $N_{\rm c} = 8$ and $10$, but only observed gas or glassy state for $N_{\rm c} = 6$ with no crystal nucleation in simulations. The dependence of phase diagram on $N_{\rm c}$ will be explored in future work.

We next consider the tetravalent PP assembly. One of the most popular models for the study of tetravalent systems is the Kern-Frenkel (KF) potential~\cite{kern2003fluid}, which is a square-well potential with angular dependence (see Supp. Mat.). The thermodynamic and crystallizability of the model have been well characterized~\cite{romano2011crystallization,rovigatti2018simulate}, and have highlighted the difficulty of obtaining nucleations of a pure crystal due to the many kinetic traps represented by competing structures.
Due to the discontinuous nature of the potential, the Monte Carlo (MC) method is commonly employed to study its assembly~\cite{miller_exploiting_2010,rovigatti2018simulate,Noya_Zubieta_Pine_Sciortino_2019}. In the following, we hence adopt MC to identify alternative stable or metastable structures competing with the desired target tetravalent lattice.


We seek the assembly of the cubic diamond (DC) lattice, probably the most sought-after crystal for photonic applications~\cite{ho1990existence}. Systems that can assemble DC lattice are almost inevitably found to be also able to assemble into hexagonal diamond (HD) lattice~\cite{romano2011crystallization}, resulting in imperfect crystals with defects and stacking faults. We adopted a tetrahedral PP design ($N_{\rm p} = 4$), and we looked for solutions that color an 8-particle unit cell of DC but cannot color a 8-particle unit cell of HD. Even in this case, our SAT solver showed that any PP solution that satisfies 8-particle DC cell can also assemble a 32-particle HD unit cell. A strategy to avoid the HD lattice in this case is to employ a larger unit cell for the DC. We hence used a larger 16-particle unit cell of a DC lattice and scanned a range of combinations of $N_{\rm s} > 8$ and $N_{\rm c}$. For each solution that we obtained for a given $N_s$ and $N_c$, we checked with the SAT solver if it can assemble into a 32-particle HD unit cell. The solution with the smallest $N_{\rm s}$ that we found to be able to form DC and not form 32-particle HD cell had $N_{\rm s} = 9$ and $N_{\rm c} = 31$, and it successfully nucleated the DC lattice in a Kern-Frenkel PP~\cite{romano2011crystallization} simulation (Fig.~\ref{fig_simulation}b), which was done with $495$ particles split equally into $9$ PP species, at number density $0.2$ (corresponding to volume fraction $0.1$). We have also carried out simulations at $0.1$ number density with $2048$ particles, which also showed homogeneous nucleation of a DC lattice (shown in the Supp. Mat.).

As the last example, we used our approach to find a system able to nucleate into the Si34 clathrate (CSi34) starting from tetrahedral PPs in the KF model. 
The smallest $N_{\rm s}$ for which a solution was found that could form CSi34 and not DC or HD lattices had $N_{\rm s} = 4$, $N_{\rm c} = 12$, and was confirmed to successfully nucleate CSi34 (Fig.~\ref{fig_simulation}c) in a simulation at $0.2$ number density with $476$ particles in species ratio $6:3:2:6$, as well as in a larger simulation at number density $0.1$ and $1904$ particles (shown in the Supp.~Mat.). 

The patch coloring and interaction matrices for all PP solutions are given in the Supp.~Mat.
While we focused so far only on designing systems for the assembly of difficult 3D lattices, our approach can be generalized to other systems, such as finite-size clusters. Our method is not limited to spherical PPs and can be used for any model where simulations or other stochastic methods can identify undesired assemblies as a list of interactions between PP species and their patches. 
It can be also combined with other techniques, such as using different strengths of interactions to disfavor undesired assemblies identified in the simulations. The approach proposed here is extensible and the systems designed in this work should be amenable of experimental realization. 

\begin{acknowledgments}
JR acknowledges support from the European Research Council Grant DLV-759187. P\v{S} acknowledges support from the ONR Grant N000142012094. JR and P\v{S} acknowledge support from the Universit{\`a} Ca' Foscari for a Visiting Scholarship.
\end{acknowledgments}

\bibliography{biblio}

\begin{thebibliography}{53}%
\makeatletter
\providecommand \@ifxundefined [1]{%
 \@ifx{#1\undefined}
}%
\providecommand \@ifnum [1]{%
 \ifnum #1\expandafter \@firstoftwo
 \else \expandafter \@secondoftwo
 \fi
}%
\providecommand \@ifx [1]{%
 \ifx #1\expandafter \@firstoftwo
 \else \expandafter \@secondoftwo
 \fi
}%
\providecommand \natexlab [1]{#1}%
\providecommand \enquote  [1]{``#1''}%
\providecommand \bibnamefont  [1]{#1}%
\providecommand \bibfnamefont [1]{#1}%
\providecommand \citenamefont [1]{#1}%
\providecommand \href@noop [0]{\@secondoftwo}%
\providecommand \href [0]{\begingroup \@sanitize@url \@href}%
\providecommand \@href[1]{\@@startlink{#1}\@@href}%
\providecommand \@@href[1]{\endgroup#1\@@endlink}%
\providecommand \@sanitize@url [0]{\catcode `\\12\catcode `\$12\catcode
  `\&12\catcode `\#12\catcode `\^12\catcode `\_12\catcode `\%12\relax}%
\providecommand \@@startlink[1]{}%
\providecommand \@@endlink[0]{}%
\providecommand \url  [0]{\begingroup\@sanitize@url \@url }%
\providecommand \@url [1]{\endgroup\@href {#1}{\urlprefix }}%
\providecommand \urlprefix  [0]{URL }%
\providecommand \Eprint [0]{\href }%
\providecommand \doibase [0]{http://dx.doi.org/}%
\providecommand \selectlanguage [0]{\@gobble}%
\providecommand \bibinfo  [0]{\@secondoftwo}%
\providecommand \bibfield  [0]{\@secondoftwo}%
\providecommand \translation [1]{[#1]}%
\providecommand \BibitemOpen [0]{}%
\providecommand \bibitemStop [0]{}%
\providecommand \bibitemNoStop [0]{.\EOS\space}%
\providecommand \EOS [0]{\spacefactor3000\relax}%
\providecommand \BibitemShut  [1]{\csname bibitem#1\endcsname}%
\let\auto@bib@innerbib\@empty
\bibitem [{\citenamefont {Whitelam}\ and\ \citenamefont
  {Jack}(2015)}]{whitelam2015statistical}%
  \BibitemOpen
  \bibfield  {author} {\bibinfo {author} {\bibfnamefont {S.}~\bibnamefont
  {Whitelam}}\ and\ \bibinfo {author} {\bibfnamefont {R.~L.}\ \bibnamefont
  {Jack}},\ }\href@noop {} {\bibfield  {journal} {\bibinfo  {journal} {Annual
  review of physical chemistry}\ }\textbf {\bibinfo {volume} {66}},\ \bibinfo
  {pages} {143} (\bibinfo {year} {2015})}\BibitemShut {NoStop}%
\bibitem [{\citenamefont {Chen}\ \emph {et~al.}(2011)\citenamefont {Chen},
  \citenamefont {Bae},\ and\ \citenamefont {Granick}}]{Chen2011}%
  \BibitemOpen
  \bibfield  {author} {\bibinfo {author} {\bibfnamefont {Q.}~\bibnamefont
  {Chen}}, \bibinfo {author} {\bibfnamefont {S.~C.}\ \bibnamefont {Bae}}, \
  and\ \bibinfo {author} {\bibfnamefont {S.}~\bibnamefont {Granick}},\ }\href
  {\doibase 10.1038/nature09713} {\bibfield  {journal} {\bibinfo  {journal}
  {Nature}\ }\textbf {\bibinfo {volume} {469}},\ \bibinfo {pages} {381}
  (\bibinfo {year} {2011})}\BibitemShut {NoStop}%
\bibitem [{\citenamefont {Chen}\ and\ \citenamefont
  {Seeman}(1991)}]{chen1991synthesis}%
  \BibitemOpen
  \bibfield  {author} {\bibinfo {author} {\bibfnamefont {J.}~\bibnamefont
  {Chen}}\ and\ \bibinfo {author} {\bibfnamefont {N.~C.}\ \bibnamefont
  {Seeman}},\ }\href@noop {} {\bibfield  {journal} {\bibinfo  {journal}
  {Nature}\ }\textbf {\bibinfo {volume} {350}},\ \bibinfo {pages} {631}
  (\bibinfo {year} {1991})}\BibitemShut {NoStop}%
\bibitem [{\citenamefont {Douglas}\ \emph {et~al.}(2009)\citenamefont
  {Douglas}, \citenamefont {Dietz}, \citenamefont {Liedl}, \citenamefont
  {H{\"o}gberg}, \citenamefont {Graf},\ and\ \citenamefont
  {Shih}}]{douglas2009self}%
  \BibitemOpen
  \bibfield  {author} {\bibinfo {author} {\bibfnamefont {S.~M.}\ \bibnamefont
  {Douglas}}, \bibinfo {author} {\bibfnamefont {H.}~\bibnamefont {Dietz}},
  \bibinfo {author} {\bibfnamefont {T.}~\bibnamefont {Liedl}}, \bibinfo
  {author} {\bibfnamefont {B.}~\bibnamefont {H{\"o}gberg}}, \bibinfo {author}
  {\bibfnamefont {F.}~\bibnamefont {Graf}}, \ and\ \bibinfo {author}
  {\bibfnamefont {W.~M.}\ \bibnamefont {Shih}},\ }\href@noop {} {\bibfield
  {journal} {\bibinfo  {journal} {Nature}\ }\textbf {\bibinfo {volume} {459}},\
  \bibinfo {pages} {414} (\bibinfo {year} {2009})}\BibitemShut {NoStop}%
\bibitem [{\citenamefont {Takeda}\ \emph {et~al.}(1999)\citenamefont {Takeda},
  \citenamefont {Umemoto}, \citenamefont {Yamaguchi},\ and\ \citenamefont
  {Fujita}}]{takeda1999nanometre}%
  \BibitemOpen
  \bibfield  {author} {\bibinfo {author} {\bibfnamefont {N.}~\bibnamefont
  {Takeda}}, \bibinfo {author} {\bibfnamefont {K.}~\bibnamefont {Umemoto}},
  \bibinfo {author} {\bibfnamefont {K.}~\bibnamefont {Yamaguchi}}, \ and\
  \bibinfo {author} {\bibfnamefont {M.}~\bibnamefont {Fujita}},\ }\href@noop {}
  {\bibfield  {journal} {\bibinfo  {journal} {Nature}\ }\textbf {\bibinfo
  {volume} {398}},\ \bibinfo {pages} {794} (\bibinfo {year}
  {1999})}\BibitemShut {NoStop}%
\bibitem [{\citenamefont {Bhatia}\ \emph {et~al.}(2009)\citenamefont {Bhatia},
  \citenamefont {Mehtab}, \citenamefont {Krishnan}, \citenamefont {Indi},
  \citenamefont {Basu},\ and\ \citenamefont
  {Krishnan}}]{bhatia2009icosahedral}%
  \BibitemOpen
  \bibfield  {author} {\bibinfo {author} {\bibfnamefont {D.}~\bibnamefont
  {Bhatia}}, \bibinfo {author} {\bibfnamefont {S.}~\bibnamefont {Mehtab}},
  \bibinfo {author} {\bibfnamefont {R.}~\bibnamefont {Krishnan}}, \bibinfo
  {author} {\bibfnamefont {S.~S.}\ \bibnamefont {Indi}}, \bibinfo {author}
  {\bibfnamefont {A.}~\bibnamefont {Basu}}, \ and\ \bibinfo {author}
  {\bibfnamefont {Y.}~\bibnamefont {Krishnan}},\ }\href@noop {} {\bibfield
  {journal} {\bibinfo  {journal} {Angewandte Chemie International Edition}\
  }\textbf {\bibinfo {volume} {48}},\ \bibinfo {pages} {4134} (\bibinfo {year}
  {2009})}\BibitemShut {NoStop}%
\bibitem [{\citenamefont {Liu}\ \emph {et~al.}(2011)\citenamefont {Liu},
  \citenamefont {Hu}, \citenamefont {Comotti},\ and\ \citenamefont
  {Ward}}]{liu2011supramolecular}%
  \BibitemOpen
  \bibfield  {author} {\bibinfo {author} {\bibfnamefont {Y.}~\bibnamefont
  {Liu}}, \bibinfo {author} {\bibfnamefont {C.}~\bibnamefont {Hu}}, \bibinfo
  {author} {\bibfnamefont {A.}~\bibnamefont {Comotti}}, \ and\ \bibinfo
  {author} {\bibfnamefont {M.~D.}\ \bibnamefont {Ward}},\ }\href@noop {}
  {\bibfield  {journal} {\bibinfo  {journal} {Science}\ }\textbf {\bibinfo
  {volume} {333}},\ \bibinfo {pages} {436} (\bibinfo {year}
  {2011})}\BibitemShut {NoStop}%
\bibitem [{\citenamefont {Rothemund}(2006)}]{Rothemund2006}%
  \BibitemOpen
  \bibfield  {author} {\bibinfo {author} {\bibfnamefont {P.~W.~K.}\
  \bibnamefont {Rothemund}},\ }\href {\doibase 10.1038/nature04586} {\bibfield
  {journal} {\bibinfo  {journal} {Nature}\ }\textbf {\bibinfo {volume} {440}},\
  \bibinfo {pages} {297} (\bibinfo {year} {2006})}\BibitemShut {NoStop}%
\bibitem [{\citenamefont {Liu}\ \emph {et~al.}(2016)\citenamefont {Liu},
  \citenamefont {Tagawa}, \citenamefont {Xin}, \citenamefont {Wang},
  \citenamefont {Emamy}, \citenamefont {Li}, \citenamefont {Yager},
  \citenamefont {Starr}, \citenamefont {Tkachenko},\ and\ \citenamefont
  {Gang}}]{liu2016diamond}%
  \BibitemOpen
  \bibfield  {author} {\bibinfo {author} {\bibfnamefont {W.}~\bibnamefont
  {Liu}}, \bibinfo {author} {\bibfnamefont {M.}~\bibnamefont {Tagawa}},
  \bibinfo {author} {\bibfnamefont {H.~L.}\ \bibnamefont {Xin}}, \bibinfo
  {author} {\bibfnamefont {T.}~\bibnamefont {Wang}}, \bibinfo {author}
  {\bibfnamefont {H.}~\bibnamefont {Emamy}}, \bibinfo {author} {\bibfnamefont
  {H.}~\bibnamefont {Li}}, \bibinfo {author} {\bibfnamefont {K.~G.}\
  \bibnamefont {Yager}}, \bibinfo {author} {\bibfnamefont {F.~W.}\ \bibnamefont
  {Starr}}, \bibinfo {author} {\bibfnamefont {A.~V.}\ \bibnamefont
  {Tkachenko}}, \ and\ \bibinfo {author} {\bibfnamefont {O.}~\bibnamefont
  {Gang}},\ }\href@noop {} {\bibfield  {journal} {\bibinfo  {journal}
  {Science}\ }\textbf {\bibinfo {volume} {351}},\ \bibinfo {pages} {582}
  (\bibinfo {year} {2016})}\BibitemShut {NoStop}%
\bibitem [{\citenamefont {Zhang}\ \emph {et~al.}(2018)\citenamefont {Zhang},
  \citenamefont {Hartl}, \citenamefont {Frank}, \citenamefont
  {Heuer-Jungemann}, \citenamefont {Fischer}, \citenamefont {Nickels},
  \citenamefont {Nickel},\ and\ \citenamefont {Liedl}}]{zhang20183d}%
  \BibitemOpen
  \bibfield  {author} {\bibinfo {author} {\bibfnamefont {T.}~\bibnamefont
  {Zhang}}, \bibinfo {author} {\bibfnamefont {C.}~\bibnamefont {Hartl}},
  \bibinfo {author} {\bibfnamefont {K.}~\bibnamefont {Frank}}, \bibinfo
  {author} {\bibfnamefont {A.}~\bibnamefont {Heuer-Jungemann}}, \bibinfo
  {author} {\bibfnamefont {S.}~\bibnamefont {Fischer}}, \bibinfo {author}
  {\bibfnamefont {P.~C.}\ \bibnamefont {Nickels}}, \bibinfo {author}
  {\bibfnamefont {B.}~\bibnamefont {Nickel}}, \ and\ \bibinfo {author}
  {\bibfnamefont {T.}~\bibnamefont {Liedl}},\ }\href@noop {} {\bibfield
  {journal} {\bibinfo  {journal} {Advanced Materials}\ }\textbf {\bibinfo
  {volume} {30}},\ \bibinfo {pages} {1800273} (\bibinfo {year}
  {2018})}\BibitemShut {NoStop}%
\bibitem [{\citenamefont {Wang}\ \emph {et~al.}(2017)\citenamefont {Wang},
  \citenamefont {Jenkins}, \citenamefont {McGinley}, \citenamefont {Sinno},\
  and\ \citenamefont {Crocker}}]{wang2017colloidal}%
  \BibitemOpen
  \bibfield  {author} {\bibinfo {author} {\bibfnamefont {Y.}~\bibnamefont
  {Wang}}, \bibinfo {author} {\bibfnamefont {I.~C.}\ \bibnamefont {Jenkins}},
  \bibinfo {author} {\bibfnamefont {J.~T.}\ \bibnamefont {McGinley}}, \bibinfo
  {author} {\bibfnamefont {T.}~\bibnamefont {Sinno}}, \ and\ \bibinfo {author}
  {\bibfnamefont {J.~C.}\ \bibnamefont {Crocker}},\ }\href@noop {} {\bibfield
  {journal} {\bibinfo  {journal} {Nature communications}\ }\textbf {\bibinfo
  {volume} {8}},\ \bibinfo {pages} {14173} (\bibinfo {year}
  {2017})}\BibitemShut {NoStop}%
\bibitem [{\citenamefont {van Anders}\ \emph {et~al.}(2013)\citenamefont {van
  Anders}, \citenamefont {Ahmed}, \citenamefont {Smith}, \citenamefont
  {Engel},\ and\ \citenamefont {Glotzer}}]{van2013entropically}%
  \BibitemOpen
  \bibfield  {author} {\bibinfo {author} {\bibfnamefont {G.}~\bibnamefont {van
  Anders}}, \bibinfo {author} {\bibfnamefont {N.~K.}\ \bibnamefont {Ahmed}},
  \bibinfo {author} {\bibfnamefont {R.}~\bibnamefont {Smith}}, \bibinfo
  {author} {\bibfnamefont {M.}~\bibnamefont {Engel}}, \ and\ \bibinfo {author}
  {\bibfnamefont {S.~C.}\ \bibnamefont {Glotzer}},\ }\href@noop {} {\bibfield
  {journal} {\bibinfo  {journal} {Acs Nano}\ }\textbf {\bibinfo {volume} {8}},\
  \bibinfo {pages} {931} (\bibinfo {year} {2013})}\BibitemShut {NoStop}%
\bibitem [{\citenamefont {Zhang}\ and\ \citenamefont
  {Glotzer}(2004)}]{zhang2004self}%
  \BibitemOpen
  \bibfield  {author} {\bibinfo {author} {\bibfnamefont {Z.}~\bibnamefont
  {Zhang}}\ and\ \bibinfo {author} {\bibfnamefont {S.~C.}\ \bibnamefont
  {Glotzer}},\ }\href@noop {} {\bibfield  {journal} {\bibinfo  {journal} {Nano
  Letters}\ }\textbf {\bibinfo {volume} {4}},\ \bibinfo {pages} {1407}
  (\bibinfo {year} {2004})}\BibitemShut {NoStop}%
\bibitem [{\citenamefont {Pawar}\ and\ \citenamefont
  {Kretzschmar}(2010)}]{pawar2010fabrication}%
  \BibitemOpen
  \bibfield  {author} {\bibinfo {author} {\bibfnamefont {A.~B.}\ \bibnamefont
  {Pawar}}\ and\ \bibinfo {author} {\bibfnamefont {I.}~\bibnamefont
  {Kretzschmar}},\ }\href@noop {} {\bibfield  {journal} {\bibinfo  {journal}
  {Macromolecular rapid communications}\ }\textbf {\bibinfo {volume} {31}},\
  \bibinfo {pages} {150} (\bibinfo {year} {2010})}\BibitemShut {NoStop}%
\bibitem [{\citenamefont {Bianchi}\ \emph {et~al.}(2011)\citenamefont
  {Bianchi}, \citenamefont {Blaak},\ and\ \citenamefont
  {Likos}}]{bianchi2011patchy}%
  \BibitemOpen
  \bibfield  {author} {\bibinfo {author} {\bibfnamefont {E.}~\bibnamefont
  {Bianchi}}, \bibinfo {author} {\bibfnamefont {R.}~\bibnamefont {Blaak}}, \
  and\ \bibinfo {author} {\bibfnamefont {C.~N.}\ \bibnamefont {Likos}},\
  }\href@noop {} {\bibfield  {journal} {\bibinfo  {journal} {Physical Chemistry
  Chemical Physics}\ }\textbf {\bibinfo {volume} {13}},\ \bibinfo {pages}
  {6397} (\bibinfo {year} {2011})}\BibitemShut {NoStop}%
\bibitem [{\citenamefont {Romano}\ and\ \citenamefont
  {Sciortino}(2011)}]{romano2011colloidal}%
  \BibitemOpen
  \bibfield  {author} {\bibinfo {author} {\bibfnamefont {F.}~\bibnamefont
  {Romano}}\ and\ \bibinfo {author} {\bibfnamefont {F.}~\bibnamefont
  {Sciortino}},\ }\href@noop {} {\bibfield  {journal} {\bibinfo  {journal}
  {Nature materials}\ }\textbf {\bibinfo {volume} {10}},\ \bibinfo {pages}
  {171} (\bibinfo {year} {2011})}\BibitemShut {NoStop}%
\bibitem [{\citenamefont {Suzuki}\ \emph {et~al.}(2009)\citenamefont {Suzuki},
  \citenamefont {Hosokawa},\ and\ \citenamefont
  {Maeda}}]{suzuki2009controlling}%
  \BibitemOpen
  \bibfield  {author} {\bibinfo {author} {\bibfnamefont {K.}~\bibnamefont
  {Suzuki}}, \bibinfo {author} {\bibfnamefont {K.}~\bibnamefont {Hosokawa}}, \
  and\ \bibinfo {author} {\bibfnamefont {M.}~\bibnamefont {Maeda}},\
  }\href@noop {} {\bibfield  {journal} {\bibinfo  {journal} {Journal of the
  American Chemical Society}\ }\textbf {\bibinfo {volume} {131}},\ \bibinfo
  {pages} {7518} (\bibinfo {year} {2009})}\BibitemShut {NoStop}%
\bibitem [{\citenamefont {Kim}\ \emph {et~al.}(2011)\citenamefont {Kim},
  \citenamefont {Kim},\ and\ \citenamefont {Deaton}}]{kim2011dna}%
  \BibitemOpen
  \bibfield  {author} {\bibinfo {author} {\bibfnamefont {J.-W.}\ \bibnamefont
  {Kim}}, \bibinfo {author} {\bibfnamefont {J.-H.}\ \bibnamefont {Kim}}, \ and\
  \bibinfo {author} {\bibfnamefont {R.}~\bibnamefont {Deaton}},\ }\href@noop {}
  {\bibfield  {journal} {\bibinfo  {journal} {Angewandte Chemie International
  Edition}\ }\textbf {\bibinfo {volume} {50}},\ \bibinfo {pages} {9185}
  (\bibinfo {year} {2011})}\BibitemShut {NoStop}%
\bibitem [{\citenamefont {Wang}\ \emph {et~al.}(2012)\citenamefont {Wang},
  \citenamefont {Wang}, \citenamefont {Breed}, \citenamefont {Manoharan},
  \citenamefont {Feng}, \citenamefont {Hollingsworth}, \citenamefont {Weck},\
  and\ \citenamefont {Pine}}]{wang2012colloids}%
  \BibitemOpen
  \bibfield  {author} {\bibinfo {author} {\bibfnamefont {Y.}~\bibnamefont
  {Wang}}, \bibinfo {author} {\bibfnamefont {Y.}~\bibnamefont {Wang}}, \bibinfo
  {author} {\bibfnamefont {D.~R.}\ \bibnamefont {Breed}}, \bibinfo {author}
  {\bibfnamefont {V.~N.}\ \bibnamefont {Manoharan}}, \bibinfo {author}
  {\bibfnamefont {L.}~\bibnamefont {Feng}}, \bibinfo {author} {\bibfnamefont
  {A.~D.}\ \bibnamefont {Hollingsworth}}, \bibinfo {author} {\bibfnamefont
  {M.}~\bibnamefont {Weck}}, \ and\ \bibinfo {author} {\bibfnamefont {D.~J.}\
  \bibnamefont {Pine}},\ }\href@noop {} {\bibfield  {journal} {\bibinfo
  {journal} {Nature}\ }\textbf {\bibinfo {volume} {491}},\ \bibinfo {pages}
  {51} (\bibinfo {year} {2012})}\BibitemShut {NoStop}%
\bibitem [{\citenamefont {Feng}\ \emph {et~al.}(2013)\citenamefont {Feng},
  \citenamefont {Dreyfus}, \citenamefont {Sha}, \citenamefont {Seeman},\ and\
  \citenamefont {Chaikin}}]{feng2013dna}%
  \BibitemOpen
  \bibfield  {author} {\bibinfo {author} {\bibfnamefont {L.}~\bibnamefont
  {Feng}}, \bibinfo {author} {\bibfnamefont {R.}~\bibnamefont {Dreyfus}},
  \bibinfo {author} {\bibfnamefont {R.}~\bibnamefont {Sha}}, \bibinfo {author}
  {\bibfnamefont {N.~C.}\ \bibnamefont {Seeman}}, \ and\ \bibinfo {author}
  {\bibfnamefont {P.~M.}\ \bibnamefont {Chaikin}},\ }\href@noop {} {\bibfield
  {journal} {\bibinfo  {journal} {Advanced Materials}\ }\textbf {\bibinfo
  {volume} {25}},\ \bibinfo {pages} {2779} (\bibinfo {year}
  {2013})}\BibitemShut {NoStop}%
\bibitem [{\citenamefont {Rechtsman}\ \emph {et~al.}(2005)\citenamefont
  {Rechtsman}, \citenamefont {Stillinger},\ and\ \citenamefont
  {Torquato}}]{rechtsman2005optimized}%
  \BibitemOpen
  \bibfield  {author} {\bibinfo {author} {\bibfnamefont {M.~C.}\ \bibnamefont
  {Rechtsman}}, \bibinfo {author} {\bibfnamefont {F.~H.}\ \bibnamefont
  {Stillinger}}, \ and\ \bibinfo {author} {\bibfnamefont {S.}~\bibnamefont
  {Torquato}},\ }\href@noop {} {\bibfield  {journal} {\bibinfo  {journal}
  {Physical review letters}\ }\textbf {\bibinfo {volume} {95}},\ \bibinfo
  {pages} {228301} (\bibinfo {year} {2005})}\BibitemShut {NoStop}%
\bibitem [{\citenamefont {Marcotte}\ \emph {et~al.}(2011)\citenamefont
  {Marcotte}, \citenamefont {Stillinger},\ and\ \citenamefont
  {Torquato}}]{marcotte2011optimized}%
  \BibitemOpen
  \bibfield  {author} {\bibinfo {author} {\bibfnamefont {E.}~\bibnamefont
  {Marcotte}}, \bibinfo {author} {\bibfnamefont {F.~H.}\ \bibnamefont
  {Stillinger}}, \ and\ \bibinfo {author} {\bibfnamefont {S.}~\bibnamefont
  {Torquato}},\ }\href@noop {} {\bibfield  {journal} {\bibinfo  {journal} {Soft
  Matter}\ }\textbf {\bibinfo {volume} {7}},\ \bibinfo {pages} {2332} (\bibinfo
  {year} {2011})}\BibitemShut {NoStop}%
\bibitem [{\citenamefont {Marcotte}\ \emph {et~al.}(2013)\citenamefont
  {Marcotte}, \citenamefont {Stillinger},\ and\ \citenamefont
  {Torquato}}]{marcotte2013designeddiamond}%
  \BibitemOpen
  \bibfield  {author} {\bibinfo {author} {\bibfnamefont {E.}~\bibnamefont
  {Marcotte}}, \bibinfo {author} {\bibfnamefont {F.~H.}\ \bibnamefont
  {Stillinger}}, \ and\ \bibinfo {author} {\bibfnamefont {S.}~\bibnamefont
  {Torquato}},\ }\href {\doibase 10.1063/1.4790634} {\bibfield  {journal}
  {\bibinfo  {journal} {The Journal of Chemical Physics}\ }\textbf {\bibinfo
  {volume} {138}},\ \bibinfo {pages} {061101} (\bibinfo {year}
  {2013})}\BibitemShut {NoStop}%
\bibitem [{\citenamefont {Zhang}\ \emph {et~al.}(2013)\citenamefont {Zhang},
  \citenamefont {Stillinger},\ and\ \citenamefont
  {Torquato}}]{zhang2013probing}%
  \BibitemOpen
  \bibfield  {author} {\bibinfo {author} {\bibfnamefont {G.}~\bibnamefont
  {Zhang}}, \bibinfo {author} {\bibfnamefont {F.}~\bibnamefont {Stillinger}}, \
  and\ \bibinfo {author} {\bibfnamefont {S.}~\bibnamefont {Torquato}},\
  }\href@noop {} {\bibfield  {journal} {\bibinfo  {journal} {Physical Review
  E}\ }\textbf {\bibinfo {volume} {88}},\ \bibinfo {pages} {042309} (\bibinfo
  {year} {2013})}\BibitemShut {NoStop}%
\bibitem [{\citenamefont {Miskin}\ \emph {et~al.}(2016)\citenamefont {Miskin},
  \citenamefont {Khaira}, \citenamefont {de~Pablo},\ and\ \citenamefont
  {Jaeger}}]{miskin2016turning}%
  \BibitemOpen
  \bibfield  {author} {\bibinfo {author} {\bibfnamefont {M.~Z.}\ \bibnamefont
  {Miskin}}, \bibinfo {author} {\bibfnamefont {G.}~\bibnamefont {Khaira}},
  \bibinfo {author} {\bibfnamefont {J.~J.}\ \bibnamefont {de~Pablo}}, \ and\
  \bibinfo {author} {\bibfnamefont {H.~M.}\ \bibnamefont {Jaeger}},\
  }\href@noop {} {\bibfield  {journal} {\bibinfo  {journal} {Proceedings of the
  National Academy of Sciences}\ }\textbf {\bibinfo {volume} {113}},\ \bibinfo
  {pages} {34} (\bibinfo {year} {2016})}\BibitemShut {NoStop}%
\bibitem [{\citenamefont {Chen}\ \emph {et~al.}(2018)\citenamefont {Chen},
  \citenamefont {Zhang},\ and\ \citenamefont {Torquato}}]{chen2018inverse}%
  \BibitemOpen
  \bibfield  {author} {\bibinfo {author} {\bibfnamefont {D.}~\bibnamefont
  {Chen}}, \bibinfo {author} {\bibfnamefont {G.}~\bibnamefont {Zhang}}, \ and\
  \bibinfo {author} {\bibfnamefont {S.}~\bibnamefont {Torquato}},\ }\href@noop
  {} {\bibfield  {journal} {\bibinfo  {journal} {The Journal of Physical
  Chemistry B}\ }\textbf {\bibinfo {volume} {122}},\ \bibinfo {pages} {8462}
  (\bibinfo {year} {2018})}\BibitemShut {NoStop}%
\bibitem [{\citenamefont {Kumar}\ \emph {et~al.}(2019)\citenamefont {Kumar},
  \citenamefont {Coli}, \citenamefont {Dijkstra},\ and\ \citenamefont
  {Sastry}}]{kumar2019inverse}%
  \BibitemOpen
  \bibfield  {author} {\bibinfo {author} {\bibfnamefont {R.}~\bibnamefont
  {Kumar}}, \bibinfo {author} {\bibfnamefont {G.~M.}\ \bibnamefont {Coli}},
  \bibinfo {author} {\bibfnamefont {M.}~\bibnamefont {Dijkstra}}, \ and\
  \bibinfo {author} {\bibfnamefont {S.}~\bibnamefont {Sastry}},\ }\href@noop {}
  {\bibfield  {journal} {\bibinfo  {journal} {arXiv preprint arXiv:1905.11061}\
  } (\bibinfo {year} {2019})}\BibitemShut {NoStop}%
\bibitem [{\citenamefont {Ducrot}\ \emph {et~al.}(2017)\citenamefont {Ducrot},
  \citenamefont {He}, \citenamefont {Yi},\ and\ \citenamefont
  {Pine}}]{ducrot2017colloidal}%
  \BibitemOpen
  \bibfield  {author} {\bibinfo {author} {\bibfnamefont {{\'E}.}~\bibnamefont
  {Ducrot}}, \bibinfo {author} {\bibfnamefont {M.}~\bibnamefont {He}}, \bibinfo
  {author} {\bibfnamefont {G.-R.}\ \bibnamefont {Yi}}, \ and\ \bibinfo {author}
  {\bibfnamefont {D.~J.}\ \bibnamefont {Pine}},\ }\href@noop {} {\bibfield
  {journal} {\bibinfo  {journal} {Nature materials}\ }\textbf {\bibinfo
  {volume} {16}},\ \bibinfo {pages} {652} (\bibinfo {year} {2017})}\BibitemShut
  {NoStop}%
\bibitem [{\citenamefont {Nelson}(2002)}]{nelson2002toward}%
  \BibitemOpen
  \bibfield  {author} {\bibinfo {author} {\bibfnamefont {D.~R.}\ \bibnamefont
  {Nelson}},\ }\href@noop {} {\bibfield  {journal} {\bibinfo  {journal} {Nano
  Letters}\ }\textbf {\bibinfo {volume} {2}},\ \bibinfo {pages} {1125}
  (\bibinfo {year} {2002})}\BibitemShut {NoStop}%
\bibitem [{\citenamefont {Manoharan}\ \emph {et~al.}(2003)\citenamefont
  {Manoharan}, \citenamefont {Elsesser},\ and\ \citenamefont
  {Pine}}]{manoharan2003dense}%
  \BibitemOpen
  \bibfield  {author} {\bibinfo {author} {\bibfnamefont {V.~N.}\ \bibnamefont
  {Manoharan}}, \bibinfo {author} {\bibfnamefont {M.~T.}\ \bibnamefont
  {Elsesser}}, \ and\ \bibinfo {author} {\bibfnamefont {D.~J.}\ \bibnamefont
  {Pine}},\ }\href@noop {} {\bibfield  {journal} {\bibinfo  {journal}
  {Science}\ }\textbf {\bibinfo {volume} {301}},\ \bibinfo {pages} {483}
  (\bibinfo {year} {2003})}\BibitemShut {NoStop}%
\bibitem [{\citenamefont {Zhang}\ \emph {et~al.}(2005)\citenamefont {Zhang},
  \citenamefont {Keys}, \citenamefont {Chen},\ and\ \citenamefont
  {Glotzer}}]{zhang2005self}%
  \BibitemOpen
  \bibfield  {author} {\bibinfo {author} {\bibfnamefont {Z.}~\bibnamefont
  {Zhang}}, \bibinfo {author} {\bibfnamefont {A.~S.}\ \bibnamefont {Keys}},
  \bibinfo {author} {\bibfnamefont {T.}~\bibnamefont {Chen}}, \ and\ \bibinfo
  {author} {\bibfnamefont {S.~C.}\ \bibnamefont {Glotzer}},\ }\href@noop {}
  {\bibfield  {journal} {\bibinfo  {journal} {Langmuir}\ }\textbf {\bibinfo
  {volume} {21}},\ \bibinfo {pages} {11547} (\bibinfo {year}
  {2005})}\BibitemShut {NoStop}%
\bibitem [{\citenamefont {Romano}\ \emph {et~al.}(2014)\citenamefont {Romano},
  \citenamefont {Russo},\ and\ \citenamefont {Tanaka}}]{romano2014influence}%
  \BibitemOpen
  \bibfield  {author} {\bibinfo {author} {\bibfnamefont {F.}~\bibnamefont
  {Romano}}, \bibinfo {author} {\bibfnamefont {J.}~\bibnamefont {Russo}}, \
  and\ \bibinfo {author} {\bibfnamefont {H.}~\bibnamefont {Tanaka}},\
  }\href@noop {} {\bibfield  {journal} {\bibinfo  {journal} {Physical review
  letters}\ }\textbf {\bibinfo {volume} {113}},\ \bibinfo {pages} {138303}
  (\bibinfo {year} {2014})}\BibitemShut {NoStop}%
\bibitem [{\citenamefont {Halverson}\ and\ \citenamefont
  {Tkachenko}(2013)}]{halverson2013dna}%
  \BibitemOpen
  \bibfield  {author} {\bibinfo {author} {\bibfnamefont {J.~D.}\ \bibnamefont
  {Halverson}}\ and\ \bibinfo {author} {\bibfnamefont {A.~V.}\ \bibnamefont
  {Tkachenko}},\ }\href@noop {} {\bibfield  {journal} {\bibinfo  {journal}
  {Physical Review E}\ }\textbf {\bibinfo {volume} {87}},\ \bibinfo {pages}
  {062310} (\bibinfo {year} {2013})}\BibitemShut {NoStop}%
\bibitem [{\citenamefont {Romano}\ and\ \citenamefont
  {Sciortino}(2012)}]{romano2012patterning}%
  \BibitemOpen
  \bibfield  {author} {\bibinfo {author} {\bibfnamefont {F.}~\bibnamefont
  {Romano}}\ and\ \bibinfo {author} {\bibfnamefont {F.}~\bibnamefont
  {Sciortino}},\ }\href@noop {} {\bibfield  {journal} {\bibinfo  {journal}
  {Nature communications}\ }\textbf {\bibinfo {volume} {3}},\ \bibinfo {pages}
  {975} (\bibinfo {year} {2012})}\BibitemShut {NoStop}%
\bibitem [{\citenamefont {Tracey}\ \emph {et~al.}(2019)\citenamefont {Tracey},
  \citenamefont {Noya},\ and\ \citenamefont {Doye}}]{tracey2019programming}%
  \BibitemOpen
  \bibfield  {author} {\bibinfo {author} {\bibfnamefont {D.~F.}\ \bibnamefont
  {Tracey}}, \bibinfo {author} {\bibfnamefont {E.~G.}\ \bibnamefont {Noya}}, \
  and\ \bibinfo {author} {\bibfnamefont {J.~P.~K.}\ \bibnamefont {Doye}},\
  }\href@noop {} {\bibfield  {journal} {\bibinfo  {journal} {The Journal of
  Chemical Physics}\ }\textbf {\bibinfo {volume} {151}},\ \bibinfo {pages}
  {224506} (\bibinfo {year} {2019})}\BibitemShut {NoStop}%
\bibitem [{\citenamefont {Morphew}\ \emph {et~al.}(2018)\citenamefont
  {Morphew}, \citenamefont {Shaw}, \citenamefont {Avins},\ and\ \citenamefont
  {Chakrabarti}}]{morphew2018programming}%
  \BibitemOpen
  \bibfield  {author} {\bibinfo {author} {\bibfnamefont {D.}~\bibnamefont
  {Morphew}}, \bibinfo {author} {\bibfnamefont {J.}~\bibnamefont {Shaw}},
  \bibinfo {author} {\bibfnamefont {C.}~\bibnamefont {Avins}}, \ and\ \bibinfo
  {author} {\bibfnamefont {D.}~\bibnamefont {Chakrabarti}},\ }\href@noop {}
  {\bibfield  {journal} {\bibinfo  {journal} {ACS nano}\ }\textbf {\bibinfo
  {volume} {12}},\ \bibinfo {pages} {2355} (\bibinfo {year}
  {2018})}\BibitemShut {NoStop}%
\bibitem [{\citenamefont {Patra}\ and\ \citenamefont
  {Tkachenko}(2018)}]{patra2018programmable}%
  \BibitemOpen
  \bibfield  {author} {\bibinfo {author} {\bibfnamefont {N.}~\bibnamefont
  {Patra}}\ and\ \bibinfo {author} {\bibfnamefont {A.~V.}\ \bibnamefont
  {Tkachenko}},\ }\href@noop {} {\bibfield  {journal} {\bibinfo  {journal}
  {Physical Review E}\ }\textbf {\bibinfo {volume} {98}},\ \bibinfo {pages}
  {032611} (\bibinfo {year} {2018})}\BibitemShut {NoStop}%
\bibitem [{\citenamefont {Ma}\ and\ \citenamefont
  {Ferguson}(2019)}]{ma2019inverse}%
  \BibitemOpen
  \bibfield  {author} {\bibinfo {author} {\bibfnamefont {Y.}~\bibnamefont
  {Ma}}\ and\ \bibinfo {author} {\bibfnamefont {A.~L.}\ \bibnamefont
  {Ferguson}},\ }\href@noop {} {\bibfield  {journal} {\bibinfo  {journal} {Soft
  matter}\ }\textbf {\bibinfo {volume} {15}},\ \bibinfo {pages} {8808}
  (\bibinfo {year} {2019})}\BibitemShut {NoStop}%
\bibitem [{\citenamefont {Tian}\ \emph {et~al.}(2020)\citenamefont {Tian},
  \citenamefont {Lhermitte}, \citenamefont {Bai}, \citenamefont {Vo},
  \citenamefont {Xin}, \citenamefont {Li}, \citenamefont {Li}, \citenamefont
  {Fukuto}, \citenamefont {Yager}, \citenamefont {Kahn} \emph
  {et~al.}}]{tian2020ordered}%
  \BibitemOpen
  \bibfield  {author} {\bibinfo {author} {\bibfnamefont {Y.}~\bibnamefont
  {Tian}}, \bibinfo {author} {\bibfnamefont {J.~R.}\ \bibnamefont {Lhermitte}},
  \bibinfo {author} {\bibfnamefont {L.}~\bibnamefont {Bai}}, \bibinfo {author}
  {\bibfnamefont {T.}~\bibnamefont {Vo}}, \bibinfo {author} {\bibfnamefont
  {H.~L.}\ \bibnamefont {Xin}}, \bibinfo {author} {\bibfnamefont
  {H.}~\bibnamefont {Li}}, \bibinfo {author} {\bibfnamefont {R.}~\bibnamefont
  {Li}}, \bibinfo {author} {\bibfnamefont {M.}~\bibnamefont {Fukuto}}, \bibinfo
  {author} {\bibfnamefont {K.~G.}\ \bibnamefont {Yager}}, \bibinfo {author}
  {\bibfnamefont {J.~S.}\ \bibnamefont {Kahn}},  \emph {et~al.},\ }\href@noop
  {} {\bibfield  {journal} {\bibinfo  {journal} {Nature Materials}\ ,\ \bibinfo
  {pages} {1}} (\bibinfo {year} {2020})}\BibitemShut {NoStop}%
\bibitem [{\citenamefont {Een}\ and\ \citenamefont
  {Sorensson}(2005)}]{een2005minisat}%
  \BibitemOpen
  \bibfield  {author} {\bibinfo {author} {\bibfnamefont {N.}~\bibnamefont
  {Een}}\ and\ \bibinfo {author} {\bibfnamefont {N.}~\bibnamefont
  {Sorensson}},\ }in\ \href@noop {} {\emph {\bibinfo {booktitle} {Proc. SAT-05:
  8th Int. Conf. on Theory and Applications of Satisfiability Testing}}}\
  (\bibinfo {year} {2005})\ pp.\ \bibinfo {pages} {502--518}\BibitemShut
  {NoStop}%
\bibitem [{\citenamefont {Liang}\ \emph {et~al.}(2018)\citenamefont {Liang},
  \citenamefont {Oh}, \citenamefont {Mathew}, \citenamefont {Thomas},
  \citenamefont {Li},\ and\ \citenamefont {Ganesh}}]{liang2018machine}%
  \BibitemOpen
  \bibfield  {author} {\bibinfo {author} {\bibfnamefont {J.~H.}\ \bibnamefont
  {Liang}}, \bibinfo {author} {\bibfnamefont {C.}~\bibnamefont {Oh}}, \bibinfo
  {author} {\bibfnamefont {M.}~\bibnamefont {Mathew}}, \bibinfo {author}
  {\bibfnamefont {C.}~\bibnamefont {Thomas}}, \bibinfo {author} {\bibfnamefont
  {C.}~\bibnamefont {Li}}, \ and\ \bibinfo {author} {\bibfnamefont
  {V.}~\bibnamefont {Ganesh}},\ }in\ \href@noop {} {\emph {\bibinfo {booktitle}
  {International Conference on Theory and Applications of Satisfiability
  Testing}}}\ (\bibinfo {organization} {Springer},\ \bibinfo {year} {2018})\
  pp.\ \bibinfo {pages} {94--110}\BibitemShut {NoStop}%
\bibitem [{\citenamefont {Papadimitriou}(1991)}]{papadimitriou1991selecting}%
  \BibitemOpen
  \bibfield  {author} {\bibinfo {author} {\bibfnamefont {C.~H.}\ \bibnamefont
  {Papadimitriou}},\ }in\ \href@noop {} {\emph {\bibinfo {booktitle} {FOCS}}},\
  Vol.~\bibinfo {volume} {91}\ (\bibinfo {year} {1991})\ pp.\ \bibinfo {pages}
  {163--169}\BibitemShut {NoStop}%
\bibitem [{\citenamefont {J{\"a}rvisalo}\ \emph {et~al.}(2012)\citenamefont
  {J{\"a}rvisalo}, \citenamefont {Le~Berre}, \citenamefont {Roussel},\ and\
  \citenamefont {Simon}}]{jarvisalo2012international}%
  \BibitemOpen
  \bibfield  {author} {\bibinfo {author} {\bibfnamefont {M.}~\bibnamefont
  {J{\"a}rvisalo}}, \bibinfo {author} {\bibfnamefont {D.}~\bibnamefont
  {Le~Berre}}, \bibinfo {author} {\bibfnamefont {O.}~\bibnamefont {Roussel}}, \
  and\ \bibinfo {author} {\bibfnamefont {L.}~\bibnamefont {Simon}},\
  }\href@noop {} {\bibfield  {journal} {\bibinfo  {journal} {Ai Magazine}\
  }\textbf {\bibinfo {volume} {33}},\ \bibinfo {pages} {89} (\bibinfo {year}
  {2012})}\BibitemShut {NoStop}%
\bibitem [{\citenamefont {Rovigatti}\ \emph {et~al.}(2018)\citenamefont
  {Rovigatti}, \citenamefont {Russo},\ and\ \citenamefont
  {Romano}}]{rovigatti2018simulate}%
  \BibitemOpen
  \bibfield  {author} {\bibinfo {author} {\bibfnamefont {L.}~\bibnamefont
  {Rovigatti}}, \bibinfo {author} {\bibfnamefont {J.}~\bibnamefont {Russo}}, \
  and\ \bibinfo {author} {\bibfnamefont {F.}~\bibnamefont {Romano}},\
  }\href@noop {} {\bibfield  {journal} {\bibinfo  {journal} {The European
  Physical Journal E}\ }\textbf {\bibinfo {volume} {41}},\ \bibinfo {pages}
  {59} (\bibinfo {year} {2018})}\BibitemShut {NoStop}%
\bibitem [{\citenamefont {Ngo}\ \emph {et~al.}(2006)\citenamefont {Ngo},
  \citenamefont {Liddell}, \citenamefont {Ghebrebrhan},\ and\ \citenamefont
  {Joannopoulos}}]{ngo2006tetrastack}%
  \BibitemOpen
  \bibfield  {author} {\bibinfo {author} {\bibfnamefont {T.}~\bibnamefont
  {Ngo}}, \bibinfo {author} {\bibfnamefont {C.}~\bibnamefont {Liddell}},
  \bibinfo {author} {\bibfnamefont {M.}~\bibnamefont {Ghebrebrhan}}, \ and\
  \bibinfo {author} {\bibfnamefont {J.}~\bibnamefont {Joannopoulos}},\
  }\href@noop {} {\bibfield  {journal} {\bibinfo  {journal} {Applied physics
  letters}\ }\textbf {\bibinfo {volume} {88}},\ \bibinfo {pages} {241920}
  (\bibinfo {year} {2006})}\BibitemShut {NoStop}%
\bibitem [{\citenamefont {Veneziano}\ \emph {et~al.}(2016)\citenamefont
  {Veneziano}, \citenamefont {Ratanalert}, \citenamefont {Zhang}, \citenamefont
  {Zhang}, \citenamefont {Yan}, \citenamefont {Chiu},\ and\ \citenamefont
  {Bathe}}]{veneziano2016designer}%
  \BibitemOpen
  \bibfield  {author} {\bibinfo {author} {\bibfnamefont {R.}~\bibnamefont
  {Veneziano}}, \bibinfo {author} {\bibfnamefont {S.}~\bibnamefont
  {Ratanalert}}, \bibinfo {author} {\bibfnamefont {K.}~\bibnamefont {Zhang}},
  \bibinfo {author} {\bibfnamefont {F.}~\bibnamefont {Zhang}}, \bibinfo
  {author} {\bibfnamefont {H.}~\bibnamefont {Yan}}, \bibinfo {author}
  {\bibfnamefont {W.}~\bibnamefont {Chiu}}, \ and\ \bibinfo {author}
  {\bibfnamefont {M.}~\bibnamefont {Bathe}},\ }\href@noop {} {\bibfield
  {journal} {\bibinfo  {journal} {Science}\ }\textbf {\bibinfo {volume}
  {352}},\ \bibinfo {pages} {1534} (\bibinfo {year} {2016})}\BibitemShut
  {NoStop}%
\bibitem [{\citenamefont {Rovigatti}\ \emph {et~al.}(2015)\citenamefont
  {Rovigatti}, \citenamefont {{\v{S}}ulc}, \citenamefont {Reguly},\ and\
  \citenamefont {Romano}}]{rovigatti2015comparison}%
  \BibitemOpen
  \bibfield  {author} {\bibinfo {author} {\bibfnamefont {L.}~\bibnamefont
  {Rovigatti}}, \bibinfo {author} {\bibfnamefont {P.}~\bibnamefont
  {{\v{S}}ulc}}, \bibinfo {author} {\bibfnamefont {I.~Z.}\ \bibnamefont
  {Reguly}}, \ and\ \bibinfo {author} {\bibfnamefont {F.}~\bibnamefont
  {Romano}},\ }\href@noop {} {\bibfield  {journal} {\bibinfo  {journal}
  {Journal of computational chemistry}\ }\textbf {\bibinfo {volume} {36}},\
  \bibinfo {pages} {1} (\bibinfo {year} {2015})}\BibitemShut {NoStop}%
\bibitem [{\citenamefont {Kern}\ and\ \citenamefont
  {Frenkel}(2003)}]{kern2003fluid}%
  \BibitemOpen
  \bibfield  {author} {\bibinfo {author} {\bibfnamefont {N.}~\bibnamefont
  {Kern}}\ and\ \bibinfo {author} {\bibfnamefont {D.}~\bibnamefont {Frenkel}},\
  }\href@noop {} {\bibfield  {journal} {\bibinfo  {journal} {The Journal of
  chemical physics}\ }\textbf {\bibinfo {volume} {118}},\ \bibinfo {pages}
  {9882} (\bibinfo {year} {2003})}\BibitemShut {NoStop}%
\bibitem [{\citenamefont {Romano}\ \emph {et~al.}(2011)\citenamefont {Romano},
  \citenamefont {Sanz},\ and\ \citenamefont
  {Sciortino}}]{romano2011crystallization}%
  \BibitemOpen
  \bibfield  {author} {\bibinfo {author} {\bibfnamefont {F.}~\bibnamefont
  {Romano}}, \bibinfo {author} {\bibfnamefont {E.}~\bibnamefont {Sanz}}, \ and\
  \bibinfo {author} {\bibfnamefont {F.}~\bibnamefont {Sciortino}},\ }\href@noop
  {} {\bibfield  {journal} {\bibinfo  {journal} {The Journal of chemical
  physics}\ }\textbf {\bibinfo {volume} {134}},\ \bibinfo {pages} {174502}
  (\bibinfo {year} {2011})}\BibitemShut {NoStop}%
\bibitem [{\citenamefont {Miller}\ and\ \citenamefont
  {Cacciuto}(2010)}]{miller_exploiting_2010}%
  \BibitemOpen
  \bibfield  {author} {\bibinfo {author} {\bibfnamefont {W.~L.}\ \bibnamefont
  {Miller}}\ and\ \bibinfo {author} {\bibfnamefont {A.}~\bibnamefont
  {Cacciuto}},\ }\href {\doibase 10.1063/1.3524307} {\bibfield  {journal}
  {\bibinfo  {journal} {J. Chem. Phys.}\ }\textbf {\bibinfo {volume} {133}},\
  \bibinfo {pages} {234108} (\bibinfo {year} {2010})},\ \bibinfo {note}
  {publisher: American Institute of Physics}\BibitemShut {NoStop}%
\bibitem [{\citenamefont {Noya}\ \emph {et~al.}(2019)\citenamefont {Noya},
  \citenamefont {Zubieta}, \citenamefont {Pine},\ and\ \citenamefont
  {Sciortino}}]{Noya_Zubieta_Pine_Sciortino_2019}%
  \BibitemOpen
  \bibfield  {author} {\bibinfo {author} {\bibfnamefont {E.~G.}\ \bibnamefont
  {Noya}}, \bibinfo {author} {\bibfnamefont {I.}~\bibnamefont {Zubieta}},
  \bibinfo {author} {\bibfnamefont {D.~J.}\ \bibnamefont {Pine}}, \ and\
  \bibinfo {author} {\bibfnamefont {F.}~\bibnamefont {Sciortino}},\ }\href
  {\doibase 10.1063/1.5109382} {\bibfield  {journal} {\bibinfo  {journal} {The
  Journal of Chemical Physics}\ }\textbf {\bibinfo {volume} {151}},\ \bibinfo
  {pages} {094502} (\bibinfo {year} {2019})}\BibitemShut {NoStop}%
\bibitem [{\citenamefont {Ho}\ \emph {et~al.}(1990)\citenamefont {Ho},
  \citenamefont {Chan},\ and\ \citenamefont {Soukoulis}}]{ho1990existence}%
  \BibitemOpen
  \bibfield  {author} {\bibinfo {author} {\bibfnamefont {K.}~\bibnamefont
  {Ho}}, \bibinfo {author} {\bibfnamefont {C.~T.}\ \bibnamefont {Chan}}, \ and\
  \bibinfo {author} {\bibfnamefont {C.~M.}\ \bibnamefont {Soukoulis}},\
  }\href@noop {} {\bibfield  {journal} {\bibinfo  {journal} {Physical Review
  Letters}\ }\textbf {\bibinfo {volume} {65}},\ \bibinfo {pages} {3152}
  (\bibinfo {year} {1990})}\BibitemShut {NoStop}%
\bibitem [{\citenamefont {Russo}\ \emph {et~al.}(2009)\citenamefont {Russo},
  \citenamefont {Tartaglia},\ and\ \citenamefont
  {Sciortino}}]{russo2009reversible}%
  \BibitemOpen
  \bibfield  {author} {\bibinfo {author} {\bibfnamefont {J.}~\bibnamefont
  {Russo}}, \bibinfo {author} {\bibfnamefont {P.}~\bibnamefont {Tartaglia}}, \
  and\ \bibinfo {author} {\bibfnamefont {F.}~\bibnamefont {Sciortino}},\
  }\href@noop {} {\bibfield  {journal} {\bibinfo  {journal} {The Journal of
  chemical physics}\ }\textbf {\bibinfo {volume} {131}},\ \bibinfo {pages}
  {014504} (\bibinfo {year} {2009})}\BibitemShut {NoStop}%
\end{thebibliography}%

 \setcounter{figure}{0}
 \makeatletter 
 \renewcommand{\thefigure}{S\@arabic\c@figure}
 \setcounter{equation}{0}
 \renewcommand{\theequation}{S\@arabic\c@equation}
 \setcounter{table}{0}
 \renewcommand{\thetable}{S\@arabic\c@table}
  \setcounter{section}{0}
 \renewcommand{\thesection}{S\@Roman\c@section}

\onecolumngrid
\newpage

\section{Supplemental Material}
\subsection{Patchy Particle Simulations}
We introduce here the patchy particle (PP) models that were used to carry out the simulations of the lattice assemblies. Both 6-valent PP model for TS assembly and tetrahedral PP model for DC and CSi34 lattice assembly were implemented in the oxDNA simulation tool package \cite{rovigatti2015comparison} and are freely available with its source code.
The energy scale is in simulation unit energy $\epsilon^*$, and the simulation temperature reported in Fig.~2 in the main text is in reduced units $\epsilon^* / k_{\rm B}$. 
For each simulated system, we first ran few simulations at empirically chosen temperatures. We identified temperature range such that at the lowest temperature, the PPs would form a glassy state, and at the highest temperature they would remain in the gas state. We then partitioned the range into different temperatures at $0.001$ interval to run respective simulations to find optimal crystallization temperature. Each simulation was started from randomly positioned non-overlapping PPs.

\subsubsection{Patchy Particle Model for Tetrastack Assembly}   
In simulations of tetrastack (TS) lattice assembly, each particle is represented by a sphere covered by 6 patches at distance $d_{\rm p} = 0.5$ distance units (d.u.) from the center of the sphere. The positions of the patches, defined in terms of the orthonormal base associated with the patchy particle, are
\begin{eqnarray*}
\mathbf{p}_1 &=& a \left( 0,1,\xi \right)\\
\mathbf{p}_2 &=& a \left( 0,-1,\xi \right)  \\
\mathbf{p}_3 &=& a \left( \xi,0,1 \right)\\
\mathbf{p}_4 &=& a \left(  0, -1, -\xi \right)  \\
\mathbf{p}_5 &=& a \left( 0,1,-\xi \right)\\
\mathbf{p}_6 &=& a \left( -\xi, 0, -1 \right)  ,
\end{eqnarray*}
where $\xi = (1 + \sqrt{5})/2$ and $a = d_{\rm p}/ \sqrt{1 + \xi^2 }$. The position of patch $i$ in the simulation box coordinate system is given by 
\begin{equation}
\mathbf{r}_{p_i} = \mathbf{r}_{\rm cm} + {\mathrm{p}_i}_x \mathbf{e}_1 + {\mathrm{p}_i}_y \mathbf{e}_2  + {\mathrm{p}_i}_z \mathbf{e}_3 
\end{equation}
where $ \mathbf{r}_{\rm cm}$ is the center of mass of the patchy particle, and $\mathbf{e}_{1,2,3}$ are the x, y, and z orthonormal base vectors associated with the patchy particle's orientation.

The interaction potential between a pair of patches on two distinct particles is 
\begin{equation}
   \label{eq_patch}
   V_{\rm patch}(r_p) =  \begin{cases} 
                         -1.001 \delta_{ij} \exp{\left[- \left( \frac{r_p}{\alpha} \right)^{10}\right]} - C & \text{if $r_p \leq r_{\rm pmax}  $} \\
                         0 & \text{otherwise}
                       \end{cases}
\end{equation}
where $\delta_{ij}$ is 1 if patch colors $i$ and $j$ can bind and 0 otherwise, $r_p$ is the distances between a pair of patches, and $\alpha = 0.12\, {\rm d.u.}$ sets the patch width. The constant $C$ is set so that for $V_{\rm patch}(r_{pmax}) = 0$, $r_{\rm pmax} = 0.18\, \rm{d.u.}$. The patchy particles further interact through excluded volume interactions ensuring that two particles do not overlap:
\begin{equation}
   f_{\rm exc}(r,\epsilon,\sigma,r^{\star}) = \begin{cases}
   V_{\rm LJ}(r, \epsilon, \sigma) & \text{if $r < r^{\star} $},\\
   \epsilon V_{\rm smooth} (r, b, r^c) & \text{if $r^{\star} < r < r^c$},\\
   0 & \text{otherwise}.
   \end{cases} 
\end{equation}
where $r$ is the distance between the centers of the patchy particles, and $\sigma$ is set to $2R = 0.8$, twice the desired radius 
of the patchy particle. The choice of a radius smaller than $d_{\rm p}$ has been done to mimic very soft particles, where the assembly has to be driven purely by bonds without being affected by excluded volume. The repulsive potential is a piecewise function, consisting of Lennard-Jones potential function
\begin{equation} 
V_{\rm LJ}(r,  \sigma) = 8 \left[ \left(\frac{\sigma}{r}\right) ^{12} - \left(\frac{\sigma}{r}\right) ^{6} \right].
\end{equation}
that is truncated using a quadratic smoothening function
\begin{equation}
V_{\rm smooth} (x, b, x^c) = b(x^c - x)^2, 
\end{equation}
with $b$ and $x_c$ are set so that the potential is a differentiable function that is equal to $0$ after a specified cutoff distance $r^c = 0.8$. 

The patchy particle system was simulated using rigid-body Molecular Dynamics with an Andersen-like thermostat~\cite{russo2009reversible} as well as using Monte Carlo simulations. During the simulation, each patch was only able to be bound to one other patch at the time, and if the binding energy between a pair of patches, as given by Eq.~\eqref{eq_patch}, is smaller than 0, none of the patches can bind to any other patch until their pair interaction potential is again 0.

\subsubsection{Patchy Particle Model for Diamond and Clathrate Lattice Assembly}   
For the diamond cubic (DC) and clathrate Si34 (CSi34) lattice assembly, we use tetravalent patchy particles with tetrahedral patch arrangement. The positions of the patches, in the orthonormal base associated with the patchy particle, are given as
\begin{eqnarray*}
\mathbf{p}_1 &=&  R \left( \sqrt{8/9}, 0 , -1/3 \right)  \\
\mathbf{p}_2 &=& R \left( -\sqrt{2/9}, \sqrt{2/3}, -1/3  \right)  \\
\mathbf{p}_3 &=& R \left( -\sqrt{2/9}, -\sqrt{2/3}, -1/3 \right)  \\
\mathbf{p}_4 &=& R \left( 0, 0, 1 \right) ,
\end{eqnarray*}
where $R = 0.5 \, \rm{d. u. }$ is the radius of the patchy particle represented by a sphere.
We perform MC simulations of the DC and CSi34 lattice assembly. Each patchy particle is modeled as a hard sphere, with excluded volume interaction between two particles at distance $r$ defined as 
\begin{equation}
  V_{\rm hs}(r) = \begin{cases}
   \infty & \text{if $r <  2 R $},\\
   0 & \text{otherwise}.
   \end{cases} 
\end{equation}
The interaction between a pair of patches $p_i$ and $q_j$ on distinct particles $i$ and $j$ is modeled through the Kern-Frenkel interaction potential:
\begin{equation}
 V_{\rm KF}(r,\theta_p, \theta_q) = \begin{cases}
   -1 & \text{if $r < 2 R + \delta $ and $\cos \theta_p < \theta_{\rm max}$ and $\cos \theta_q < \theta_{\rm max}$},\\
   0 & \text{otherwise}.
   \end{cases} 
\end{equation}
where $\delta$ is set to $0.12$ d.u.~in our simulations and $\theta_{\rm max}$ is set to $0.98$. Furthermore, we use $\mathbf{r} = \mathbf{r}_{{\rm cm}_q} -\mathbf{r}_{{\rm cm}_p}$, where $r = \norm{\mathbf{r}}$ is the distance between the centers of mass of the patchy particles $p$ and $q$, to define angles
\begin{eqnarray}
 \cos \theta_p &=& \frac{\mathbf{r} \cdot \mathbf{p_i}}{ \norm{\mathbf{r}}  \norm{\mathbf{p_i}} } \\
 \cos \theta_q &=& \frac{\mathbf{-r} \cdot \mathbf{q_j}}{ \norm{\mathbf{r}}  \norm{\mathbf{q_j}} } 
\end{eqnarray}
where $\mathbf{p_i}$ is the vector from center of mass of particle $p$ towards patch $p_i$, and analogously for patch $q_j$.

\subsubsection{Simulations of DC and CSi34 Lattice Assembly at 0.1 Number Density}
Additionally to simulations shown in Fig.~2 in the main text, we have further simulated the homogeneous nucleation of the DC lattice with a simulation $2048$ particles (at $0.1$ number density, corresponding to $\approx 0.05$ volume fraction), split into the 9 respective species in ratio $2:2:2:1:2:2:2:1:2$, which corresponds to the number of times each PP species is represented in the 16-particle DC unit cell. We further simulated CSi34 lattice assembly with 1904 particles respectively, split into 4 species in ratio $6:3:2:6$, corresponding to the relative PP species ratio in the 34-particle unit cell of CSi34. Both systems have shown homogeneous nucleation and the summary of the simulations is in Fig.~S1.

\begin{figure*}[t]
   \centering
   \includegraphics[width=0.9\textwidth]{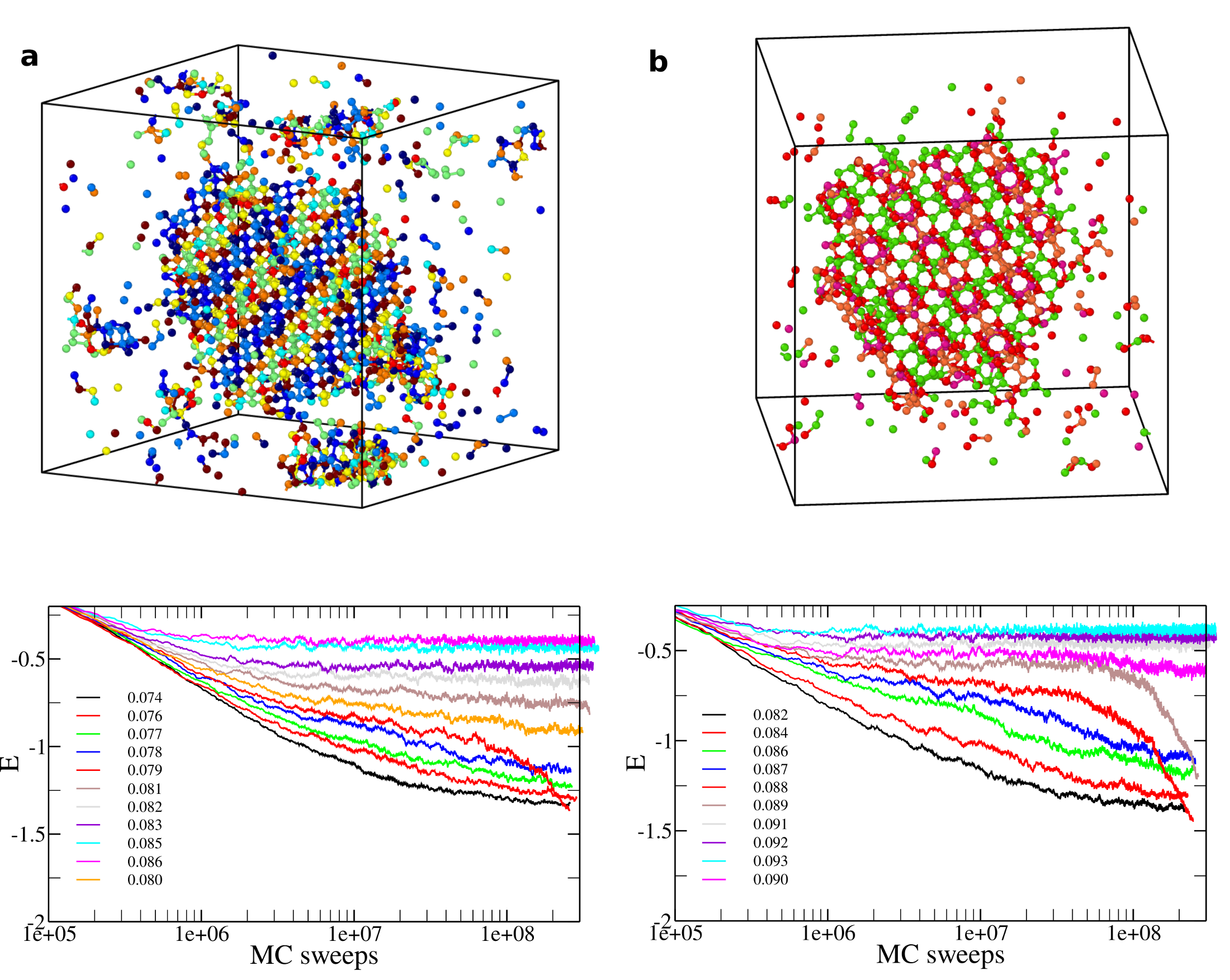}
   \caption{Further simulations of assembly of \textbf{a)} diamond cubic, \textbf{b)} clathrate Si34 lattices respectively. The system size is $N=2048$ for the DC lattice (a), and $N=1904$ for the CSi34 lattice (b). Top panels show simulations snapshots after the nucleation event. Bottom panels show the energy as a function of Monte Carlo Sweeps, displaying the nucleation events as a sudden decrease in energy of the system at the nucleation temperature.}
   \label{fig_supplementary}
\end{figure*}

\subsection{Patchy Particle Solutions for Crystal Lattices}
We list here the patch coloring and color interaction rules that were found by the SAT solver and verified by simulations to assemble into the TS, DC, and CSi34 lattices respectively. For each PP species, the patch colorings are specified as a list $\left(p, c \right)$, where $p$ is a number that identifies the patch on a particle (1 to 6 for patchy particles that assemble in TS lattice, and 1 to 4 for patchy particles that assemble in DC or CSi34 lattice), and $c$ identifies the patch color (from $1$ to $N_c$, where $N_c$ is the total number of colors used). The interacting colors are then listed as pairs $(c_i,c_j)$, where color $c_i$ can only bind to color $c_j$ and vice versa. For self-complementary colors, we list $(c_i,c_i)$. The designs of patchy particles for assembly of TS, DC, and CSi34 lattices respectively are given below:\\
\begin{itemize}
\item 
TS crystal lattice  design with $N_s = 2$ and $N_c = 12$:
\begin{center}
\begin{tabular}{|l|llllll|}
\hline
PP species & \multicolumn{6}{|c|}{Patch Coloring}\\
\hline
\mbox{1: }  &(1,1) & (2,2) & (3,3) & (4,4)  & (5,5)  & (6,6) \\
\mbox{2: }  &(1,7) & (2,8) & (3,9) & (4,10) & (5,11) & (6,12) \\
\hline
\multicolumn{7}{|c|}{Color interactions}\\
\hline
\multicolumn{7}{|l|}{(1,5), (2,12), (3,8), (4,7), (6,11), (9,10)} \\
\hline
\end{tabular}
\end{center}
%
\item DC crystal design with $N_s = 9$ and $N_c = 31$:
\begin{center}
\begin{tabular}{|l|llll|}
\hline
PP species & \multicolumn{4}{|c|}{Patch Coloring}\\
\hline
\mbox{1: } &(1,16) &(2,9)  &(3,18) &(4,4) \\
\mbox{2: } &(1,26) &(2,13) &(3,23)  &(4,1) \\
\mbox{3: } &(1,5)  &(2,8)  &(3,24) &(4,31)  \\
\mbox{4: } &(1,12) &(2,21) &(3,27)  &(4,19)\\
\mbox{5: } &(1,12) &(2,29)  &(3,14) &(4,17)\\
\mbox{6: } &(1,28) &(2,15)  &(3,7)   &(4,6)  \\
\mbox{7: } &(1,3)  &(2,22)  &(3,11)  &(4,2)  \\
\mbox{8: } &(1,21) &(2,30)  &(3,12)  &(4,19)\\
\mbox{9: } &(1,20) &(2,25)  &(3,8)   &(4,10) \\
\hline
\multicolumn{5}{|c|}{Color interactions}\\
\hline
\multicolumn{5}{|l|}{(1,4), (2,25), (11,17), (12,12), (13,13),(6,18)}\\
\multicolumn{5}{|l|}{(16,31), (19,22), (20,23), (3,9), (5,26),  (7,14)} \\
\multicolumn{5}{|l|}{(21,28), (24,24), (27,30), (29,29),  (8,8), (10,15)}\\
\hline
\end{tabular}
\end{center}
%
\item CSi34 crystal design with $N_s = 4$ and $N_c = 12$:
\begin{center}
\begin{tabular}{|l|llll|}
\hline
PP species & \multicolumn{4}{|c|}{Patch Coloring}\\
\hline
\mbox{1: }& (1,3)&(2,11)&(3,8)&(4,5) \\
\mbox{2: }& (1,12)&(2,9)&(3,4)&(4,12) \\
\mbox{3: }& (1,7)&(2,7)&(3,4)&(4,7) \\
\mbox{4: }& (1,6)&(2,10)&(3,1)&(4,2) \\
\hline
\multicolumn{5}{|c|}{Color interactions}\\
\hline
\multicolumn{5}{|l|}{ (1,1) (2,2) (3,12) (4,4) (5,6) (7,8) (9,9) (10,11)} \\
\hline
\end{tabular}
\end{center}
\end{itemize}

\subsection{Boolean Clause Formulation for SAT}
As described in the main text, we formulate the PP design problem as a SAT problem by specifying it as a set of binary clauses in a format acceptable by SAT solver software. We provide here detailed formulation of the binary clauses in a format required as an input into the commonly used SAT solvers.

The set of binary clauses as defined in Table 1 in the main text fully specify the PP design problem. 
For a target lattice structure with unit cell consisting of $L$ positions, where each position has $N_p$ neighbors (and hence $N_p$ slots), we are looking for a solution that has $N_s$ PP species and uses in total $N_c$ different colors. The particle geometry (i.e. patch positions on the PP) is hard-coded by the lattice geometry, and we need to provide all possible rotations $N_o$ that a PP can make in a given lattice position so that its patches overlap with the slots of the given position (illustrated in Fig.~1 in the main text). For the PPs used for TS lattice assembly $N_o = 6$. For tetrahedral PPs used in DC and CSi34 lattices, $N_o = 12$. For each PP orientation $o$, we assign a mapping $\phi_o$, which maps patch on the PP to the slot of the lattice position, e.g. $\phi_1 = (1,2,3,4,5,6)\rightarrow(2,3,1,5,6,4)$ for a particle with 6 patches used for TS assembly.


We define binary variables $x^{\rm int}_{c_i,c_j}$, which are 1 if given pair $c_i \leq c_j \in [1,N_{\rm c}] $ of colors can interact and 0 if they cannot. Next set of variables $x^{\rm pcol}_{s,p,c}$ define coloring of patches for each particle species $s \in  [1,N_{\rm s}]$ and $x^{\rm pcol}_{s,p,c}$ is 1 if $p$-th patch ($p \in  [1,N_{\rm p}])$ of $s$-th particle species is assigned to have color $c$. We further introduce a set of variables that define arrangement of PP types in the lattice, where $x^L_{l,s,r}$ is 1 if in the desired lattice geometry, the PP species that occupies position $l \in  [1,L]$ is of PP type $s \in [1,N_{\rm s}]$ and its orientation is set to $o \in [1,N_o]$. Lastly, we define variables $x^A_{l,k,c}$, which is 1 if the PP in lattice position $l \in [1,L]$ is oriented in such a way that the slot $k \in [1,N_{\rm p}]$ of that position overlaps with PP's patch that has color $c$. 
The variables are defined for all possible combinations of colors, PP species, patches, orientations, lattice positions.  The solution is specified by a list of variables that are 1 (true). For instance $x^{\rm int}_{c_a,c_b} = 1$ means that in the color interaction matrix, color $c_a$ is compatible with $c_b$. 
However, to make sure that the assignment of true and false values to all defined variables  is a correct solution to the design task, we need to define binary clauses that introduce relations between the variables that have to be satisfied. 

The clauses (i)-(iii) in Table 1 ensure feasibility of the solution: as we define binary variables $x^{\rm pcol}, x^L, x^{\rm int}$ corresponding to all possible combinations of color interactions, patch coloring and PP assignment to positions on the lattice, the first three sets of clauses ensure that in the obtained solution, the variables that are mutually exclusive (e.g.~a patch having at the same time two different colors) cannot be both true at the same time. The clauses (iv)-(v) enforce that for the correct solution, the PPs have to be arranged in the target lattice in such a way that patches in contact have compatible colors. Finally, clauses (vi)-(vii) impose that for the solution found by the SAT solver, all $N_{\rm s}$ PP species have to be included in the lattice formation, and all $N_{\rm c}$ colors have to be used. This requirement is added to avoid the solvers coming up with trivial solutions, such as designing one PP species with all patches colored to a self-complementary color, which would then trivially satisfy the target lattice.

As an input for the SAT solver, the problem has to be formulated in terms of clauses $C_j$, each of them containing variables $x_i$ connected by OR clauses. The final SAT problem corresponds to all respective clauses $C_j$ connected by AND clauses. To conform with this input format, the Boolean clauses introduced in Table 1 in the main text can be all reformulated as detailed below. The final SAT problem is a conjunction of all individual clauses from sets (i)-(vii) described below. The definitions use the following logic symbols: $\neg$: negation; $\land$: conjunction (AND); $\lor$: disjunction (OR); $\implies$: implies; $\iff$: if and only if.

\begin{enumerate}[(i)]
\item  Each color $c_i$ can only bind to one other color $c_j$ (including possible self-complementarity).
\begin{equation}
\label{eq_exactly_one_color}
\forall  c_i \leq c_j < c_k \in [1,N_c]: C^{\rm int}_{c_i,c_j,c_k} = \neg x^{\rm int}_{c_i,c_j} \lor \neg x^{\rm int}_{c_i,c_k} .
\end{equation}
To illustrate how the above clauses achieve unique binding between colors, consider variable $x^{\rm int}_{c_a,c_b} = 1$ for particular choice of $c_a$ and $c_b$ (meaning that these colors can bind). We consider a color $c_k$ different from $c_a$ and $c_b$. The SAT problem includes conjunction  of all clauses $C^{\rm int}$ as defined in Eqs.~\ref{eq_exactly_one_color}. Hence all of the clauses have to be true, including the clause $C^{\rm int}_{c_a,c_b,c_k} = \neg x^{\rm int}_{c_a,c_b} \lor \neg x^{\rm int}_{c_a,c_k}$. Since $\neg x^{\rm int}_{c_a,c_b} $ is false, satisfying this clause is only possible if $x^{\rm int}_{c_a,c_k} = 0$, i.e. colors $c_a$ and $c_k$ are not allowed to interact. Analogously, we can show that $x^{\rm int}_{c_b,c_k}$ must be $0$ and therefore $c_b$ and $c_k$ do not interact either.

\item Each patch $p$ of each PP species $s$ is assigned exactly one color:
\begin{equation} 
\label{eq_exactly_one_patch}
\forall s \in [1,N_s], p \in [1,N_p], c_l < c_k \in [1,N_c]: C^{\rm pcol}_{s,p,c_k,c_l} = \neg x^{pcol}_{s,p,c_k} \lor \neg x^{\rm pcol}_{s,p,c_l}.
\end{equation}
In a manner analogous to Eqs.~\ref{eq_exactly_one_color}, the clauses $C^{\rm pcol}$ ensure that if for example $x^{pcol}_{s,p,c_a}$ is 1, we need to have $x^{pcol}_{s,p,c_k} = 0$ for all other $c_k \neq c_a$ in order to satisfy the $C^{\rm pcol}$ clauses, and hence patch $p$ on PP species $s$ can only have color $c_a$.

\item Each lattice position $l$ is only assigned exactly one PP species with exactly one assigned orientation:
\begin{equation} 
\label{eq_exactly_one_pos}
\forall l \in [1,L], s_i < s_j \in [1,N_s], o_i < o_j \in [1,N_o]:  C^L_{l,s_i,o_i,s_j,o_j} = \neg x^L_{l,s_i,o_i} \lor \neg x^L_{l,s_j,o_j} .
\end{equation}
Analogously to clauses (i) and (ii), the set of clauses defined in Eqs.~\ref{eq_exactly_one_pos} ensure that there can be only one PP species with only one assigned orientation occupying given slot $l$ in the target lattice.

\item For all pairs of slots $k_i$ and $k_j$ that are in contact in neighboring lattice positions $l_i, l_j$ (e.g. as shown in Fig.~1b in main text), the patches that occupy them need to have complementary colors:
\begin{equation*}
\forall c_i \leq c_j \in [1,N_c]: C^{\rm lint}_{l_i,k_i,l_j,k_j,c_i,c_j} = \left( x^A_{l_i,k_i,c_i} \land  x^A_{l_j,k_j,c_j} \right) \implies x^C_{c_i,c_j},
\end{equation*}
which can be equivalently rewritten as 
\begin{equation}
\label{eq_slots}
C^{\rm lint}_{l_i,k_i,l_j,k_j,c_i,c_j} = \neg x^A_{l_i,k_i,c_i} \lor \neg  x^A_{l_j,k_j,c_j} \lor x^C_{c_i,c_j}.
\end{equation}
These clauses assure that PPs placed in neighboring positions in the lattice interact through the correctly colored slots. The Eqs.~\ref{eq_slots} hence encode the geometry of the target lattice. 

\item The slot of lattice position $l$ is colored with the same color as the patch of the PP species occupying it:
\begin{equation*}
\forall l \in [1,L], k \in [1,N_p], o \in [1,N_o], s \in [1,N_s], c \in [1,N_c]: C^{\rm LS}_{l,s,o,c,k} =   x^L_{l,s,o} \implies \left( x^A_{l, k, c} \iff x^{\rm pcol}_{s, \phi_o(k), c} \right) , 
\end{equation*}
which can be equivalently rewritten as 
\begin{equation}
C^{\rm LS}_{l,s,o,c,k} = \left( \neg  x^L_{l,s,o} \lor \neg x^A_{l, k, c} \lor x^{\rm pcol}_{s, \phi_o(k), c} \right) \land  \left(  \neg  x^L_{l,s,o} \lor  x^A_{l, k, c} \lor \neg x^{\rm pcol}_{s, \phi_o(k), c} \right) 
\label{eq_latice_equiv}
\end{equation}
These clauses are required to correctly set variables $x^A$, which are used in clauses (iv). 
A PP of type $s$ can be placed in $N_o$ different orientations $o$ into a specific position $l$ on the lattice. For a particular choice of $o$, the mapping function $\phi_o$ maps $\phi_o(k)$-th patch to $k$-th slot, assuring that variable $x^A_{l,k,c}$ is 1 if the $\phi_o(k)$-th patch of particle on lattice position $l$ has color $c$.  
\item All $N_s$ PP species have to be used at least once in the assembled lattice: 
\begin{equation}
\forall s \in [1,N_s]: C^{\rm all\,s.}_{s} =  \bigvee_{\forall l \in [1,L], o \in [1,N_o] } x^L_{l,s,o} 
\end{equation}
For each $s$, the variables in the clause $C^{\rm all\,s.}_{s}$ are connected by OR (disjunction). Each clause contains all combinations of variables for all lattice positions $l$ and orientations $o$. These clauses ensure that in the solution, each PP species appears at least once in the lattice. For examples, consider that there is a solution that does not contain PP species number $s_a$ in the lattice. In that case, variables $x^L_{l,s_a,o}$ are 0 for all values of $l$ and $o$, which makes the clause $C^{\rm all\,s.}_{s_a}$ false. For a valid solution to the SAT problem, however, all clauses need to be true.

\item Each color $c$ of $N_c$ total number of colors is assigned to at least one patch of one of the PP species: 
\begin{equation}
\forall c \in [1,N_c]: C^{\rm all\,c.}_{c} =  \bigvee_{\forall s \in [1,N_s], p \in [1,N_p] } x^{\rm pcol}_{s,p,c}.
\end{equation}
These clauses ensure that all colors are used in the solution in a similar way that clauses (vi) ensure that all PP species are used.
\end{enumerate}

For the design of the TS lattice, we introduced an additional set of clauses that ensure that any pair of particles (of the same or different species) cannot bind by more than one bond at a time:
$\forall s_i, s_j \in [1,N_s], c^1_i,c^2_i,c^1_j,c^2_j \in [1,N_c]:$
\begin{equation}
 C^{\rm no\,two}_{s_i,s_j,p^i_1,p^i_2,p^j_1,p^j_2,c^1_i,c^2_i,c^1_j,c^2_j} = \neg \left(  x^{\rm pcol}_{s_i,p^i_1,c^1_i} \land x^{\rm pcol}_{s_i,p^i_2,c^2_i} \land x^{\rm pcol}_{s_j,p^j_1,c^1_j}  \land x^{\rm pcol}_{s_j,p^j_2,c^2_j} \land x^{\rm int}_{c^1_i,c^1_j}\land x^{\rm int}_{c^2_i,c^2_j} \right),
\label{eq_no_more_than_two}
\end{equation}
where $p^i_1,p^i_2,p^j_1,p^j_2$ are all possible pairs of patches on PP of type $s_i$ and $s_j$ respectively for which there is a possible orientation so that they can both bind if they have compatible colors. 

To find solutions for SAT problems considered in this work, we used MiniSat, MapleSAT, or Walksat solvers \cite{een2005minisat,papadimitriou1991selecting,liang2018machine}.
These are popular standard tools used by researchers in constraint satisfaction problems community and we used them as 'black box'. We found Walksat to be the fastest in finding the solutions to most of our PP design problems (typically in several seconds). However, if the solution does not exist, Walksat algorithm is unable to prove the impossibility, as it just continues its search for a solution even if it does not exist. MapleSAT had performance similar to MiniSat in terms of time it took them to find a solution (between few seconds to tens of minutes), but MapleSAT is more memory efficient (less than $2\, {\rm GB}$ RAM for SAT problems that we encountered in this work), allowing us to run multiple MapleSAT solvers in parallel without running out of available memory on a typical workstation. As opposed to Walksat, MiniSat and MapleSAT can also be used to prove that no solution exists for a given SAT problem. However, for certain combinations of $N_{\rm s}$ and $N_{\rm c}$ for DC and CSi34 lattices, the algorithms were not able to find solution nor prove impossibility within the maximum 2 hours running time that we imposed for the solvers. It is possible the solutions could be found (or impossibility proved) if algorithms were run for longer.   

For each successful solution we found for a given lattice and $N_s$ and $N_c$, we tested the solution for its ability to assemble into undesired structures using MiniSat. In this case, we explicitly set true value to the combination of patch coloring and color interactions ($x^{\rm int}$ and $x^{\rm pcol}$) variables that encodes the solution that we found,  and use clauses (i)-(v) to formulate the SAT problem. For this  task, it takes MiniSat (or MapleSat) only few seconds to find out if the PP species (or their subset) can or cannot assemble into a given undesired lattice. Hence, we can test the solution very quickly against a list of known undesired structures.

The software implementing the conversion of the PP design to logic clause formulation in a format accepted by the SAT software tools is available upon request.

\end{document}